\documentclass[%
 reprint,
 superscriptaddress,
 twocolumn,
 amsmath,amssymb,
 aps,
prx,
floatfix,
]{revtex4-2}
\usepackage{graphicx}% Include figure files
\usepackage{dcolumn}% Align table columns on decimal point
\usepackage{bm}% bold math
\usepackage{siunitx}
\usepackage{orcidlink}

\usepackage[T1]{fontenc}

\usepackage{enumitem}

\renewcommand{\thefigure}{\,\arabic{figure}}

\usepackage{import}
\usepackage[columnwise]{lineno}
\usepackage{hyperref}% add hypertext capabilities

\newcommand{\beginsupplement}{%
        \setcounter{table}{0}
        \renewcommand{\thetable}{S\arabic{table}}%
        \setcounter{figure}{0}
        \renewcommand{\thefigure}{S\arabic{figure}}%
        \setcounter{section}{0}
        \renewcommand{\thesection}{S\arabic{section}}
     }

\begin{document}
%%%%%%%%%%%%%%%%%%%%%%%%%%%%%%%%%%%%%%%%%%%%%%%%%%%%%%%%%%%%%
\title{Automated electrostatic characterization of quantum dot devices\\ in single- and bilayer heterostructures}% Force line breaks with \\

\author{Merritt P. R. Losert}
\email{merritt.losert@nist.gov}
\affiliation{National Institute of Standards and Technology, Gaithersburg, MD 20899, USA}
\affiliation{Joint Center for Quantum Information and Computer Science, University of Maryland, College Park, MD 20742, USA}

\author{Dario Denora\orcidlink{0009-0007-1443-6613}}
\affiliation{QuTech and Kavli Institute of Nanoscience, Delft University of Technology, PO Box 5046, 2600 GA Delft, The Netherlands}

\author{Barnaby van Straaten\orcidlink{0009-0008-2113-8523}}
\affiliation{QuTech and Kavli Institute of Nanoscience, Delft University of Technology, PO Box 5046, 2600 GA Delft, The Netherlands}

\author{Michael Chan}
\affiliation{QuTech and Kavli Institute of Nanoscience, Delft University of Technology, PO Box 5046, 2600 GA Delft, The Netherlands}

\author{Stefan D. Oosterhout\orcidlink{0000-0001-9648-5206}}
\affiliation{QuTech and Netherlands Organisation for Applied Scientific Research (TNO), Delft, The Netherlands}

\author{Lucas Stehouwer\orcidlink{0009-0003-2891-4795}}
\affiliation{QuTech and Kavli Institute of Nanoscience, Delft University of Technology, PO Box 5046, 2600 GA Delft, The Netherlands}

\author{Giordano Scappucci\orcidlink{0000-0003-2512-0079}}
\affiliation{QuTech and Kavli Institute of Nanoscience, Delft University of Technology, PO Box 5046, 2600 GA Delft, The Netherlands}

\author{Menno Veldhorst\orcidlink{0000-0001-9730-3523}}
\affiliation{QuTech and Kavli Institute of Nanoscience, Delft University of Technology, PO Box 5046, 2600 GA Delft, The Netherlands}

\author{Justyna P. Zwolak\orcidlink{0000-0002-2286-3208}}
\email{jpzwolak@nist.gov}
\affiliation{National Institute of Standards and Technology, Gaithersburg, MD 20899, USA}
\affiliation{Joint Center for Quantum Information and Computer Science, University of Maryland, College Park, MD 20742, USA}
\affiliation{Department of Physics, University of Maryland, College Park, MD 20742, USA}

\date{\today}
%%%%%%%%%%%%%%%%%%%%%%%%%%%%%%%%%%%%%%%%%%%%%%%%%%%%%%%%%%%%%
\begin{abstract}
As quantum dot (QD)-based spin qubits advance toward larger, more complex device architectures, rapid, automated device characterization and data analysis tools become critical.
The orientation and spacing of transition lines in a charge stability diagram (CSD) contain a fingerprint of a QD device's capacitive environment, making these measurements useful tools for device characterization.
However, manually interpreting these features is time-consuming, error-prone, and impractical at scale.
Here, we present an automated protocol for extracting underlying capacitive properties from CSDs.
Our method integrates machine learning, image processing, and object detection to identify and track charge transitions across large datasets without manual labeling.
We demonstrate this method using experimentally measured data from a strained-germanium single-quantum-well (planar) and a strained-germanium double-quantum-well (bilayer) QD device.
Unlike for planar QD devices, CSDs in bilayer germanium heterostructure exhibit a larger set of transitions, including interlayer tunneling and distinct loading lines for the vertically stacked QDs, making them a powerful testbed for automation methods.
By analyzing the properties of many CSDs, we can statistically estimate physically relevant quantities, like relative lever arms and capacitive couplings. 
Thus, our protocol enables rapid extraction of useful, nontrivial information about QD devices.
\end{abstract}

\maketitle
%%%%%%%%%%%%%%%%%%%%%%%%%%%%%%%%%%%%%%%%%%%%%%%%%%%%%%%%%%%%%
\section{Introduction}
%%%%%%%%%%%%%%%%%%%%%%%%%%%%%%%%%%%%%%%%%%%%%%%%%%%%%%%%%%%%%
Gate-defined quantum dots (QDs) in semiconductor heterostructures have emerged as a leading platform for spin-based quantum information processing and quantum simulation, owing to their compatibility with advanced semiconductor fabrication and prospects for dense integration in two-dimensional architectures~\cite{Loss98-QCD, Zwanenburg13-SQE, Burkard21-SSQ, Borsoi22-QCA, Weinstein22-ULS, Neyens24-PQW}.
Their long coherence times and small size have enabled single- and two-qubit gate fidelities above 99~\%~\cite{Veldhorst14-AQD, Yoneda18-QDS,  Xue22-QLS, Mills22-TSP, Huang24-HFS, Watzinger18-GER, John24-TAL}.
Recent progress in materials growth, gate-stack engineering, and control electronics has enabled increasingly large arrays of QDs, demonstrated across several material platforms and device architectures~\cite{Zajac16-SGA, Mills19-SSC, Ha22-FDQ, George25-SQA, DeSmet25-HFS, Scappucci21-GER, Sammak19-SUG, Hendrickx19-GCQ}.
Among these, devices based on holes confined in strained-germanium single and double quantum wells, combining high-quality heterostructures~\cite{Sammak19-SUG, Hendrickx19-GCQ, Stehouwer23-GWS, Tidjani25-3DA} with small effective masses and strong spin–orbit coupling~\cite{Lodari21-LPD} and promising prospects for high-fidelity gate operations~\cite{Scappucci21-GER, Wang24-HOP, Tosato25-QARPET}, have attracted significant attention.

\begin{figure}[t]
    \centering
    \includegraphics[width=\columnwidth]{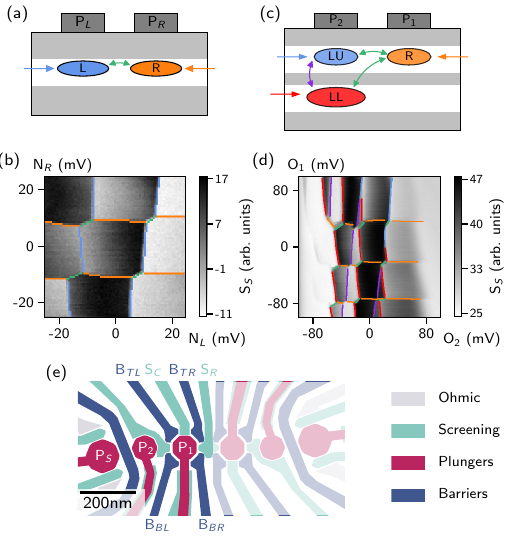}
    \caption{
    \textbf{Problem visualization for planar and bilayer QD device.}
    (a) Schematic heterostructure for a planar germanium hole device.
    (b) Sample CSD from a planar germanium device, adapted from Ref.~\cite{Rao24-MAViS}, with left (blue), right (yellow), and interdot (green) transitions highlighted. 
    This CSD is taken in the normalized plungers space. 
    (c) Schematic bilayer germanium heterostructure, with two QDs formed under gate $\mathrm{P}_2$ and one QD formed under gate $\mathrm{P}_1$. 
    (d) Example CSD taken for the bilayer system outlined in (c), with left upper (blue), left lower (red), interlayer (purple), right (yellow), and left-right interdot (green) transitions labeled. 
    This CSD is taken in the orthogonalized plungers space, where $\mathrm{O}_1$ and $\mathrm{O}_2$ are orthogonalized versions of the plungers $\mathrm{P}_1$ and $\mathrm{P}_2$ shown in (e).
    (e) A schematic illustration of the bilayer device used in this work, with the sensor $\mathrm{P}_S$, plungers $\mathrm{P}_2$ and $\mathrm{P}_1$, and other relevant gates labeled.
    The part of the device not used in experiments is grayed out.
    }
    \label{fig:intro}
\end{figure}

As QD arrays continue to scale, it is increasingly important to automate all aspects of device characterization, calibration, and control~\cite{Zwolak21-AAQ}.
The electrostatic environment of the QD device required to confine single holes is controlled by applying voltages to metallic gates fabricated on top of the heterostructure. 
To measure charge occupancy, a two-dimensional (2D) map of the sensor QD's conductance as a function of the two plunger gates is generated; this map is commonly referred to as the \textit{charge stability diagram} (CSD).
A sample CSD measured on a double-QD system in a planar germanium hole array, illustrated schematically in Fig.~\ref{fig:intro}(a), is shown in Fig.~\ref{fig:intro}(b).
Each line in the CSD corresponds to a charge transition and represents either loading or unloading a carrier into or out of a QD from a neighboring charge reservoir, or transferring a carrier between neighboring QDs.
The orientations and spacings of these transition lines encode a wealth of electrostatic information.
By measuring the angles of the different line families, one can infer the relative lever arms between gates and QDs, and by analyzing their spacings, one can extract charging energies and mutual capacitances up to an overall energy scale~\cite{Wiel02-DQD}.
Combining these quantities yields an electrostatic characterization of the device in terms of a constant-capacitance model, providing direct feedback on gate layout and heterostructure design~\cite{Wiel02-DQD, vanStraaten24-QAR, Gualtieri25-QDS}.
Such information is also critical for later stages of device tune-up, including virtual-gate construction and robust control protocols~\cite{Oakes2021-AVE, Ziegler23-AEC, Rao24-MAViS}.

While lateral scaling in planar arrays is a natural route to increasing the number of qubits~\cite{Zajac16-SGA, Mills19-SSC, Ha22-FDQ, Borsoi22-QCA, John24-TAL, George25-SQA}, the construction of three-dimensional QD arrays using multi-quantum well structures enables vertically stacked planes of qubits, as well as higher connectivity between them~\cite{Tosato22-HHB, Tidjani25-3DA}.
  
Figure~\ref{fig:intro}(c) shows a schematic of a germanium double quantum well device with two vertically stacked QDs---left upper (LU) and left lower (LL)---under plunger gate $\mathrm{P}_2$ and a third, right dot (R) under a neighboring plunger $\mathrm{P}_1$.
The corresponding CSD, shown in Fig.~\ref{fig:intro}(d), exhibits a richer pattern of loading and interdot transitions than conventional planar double-QD devices, because the QDs are confined under the same plunger but in different wells with finite tunnel coupling between them~\cite{Tosato22-HHB, Tidjani25-3DA, Ivlev24-CVD}.
While this expanded phenomenology is attractive for device design and connectivity, it also makes manual electrostatic analysis and optimal orthogonal QD control substantially more challenging.
Given the unique challenges they present, these devices serve as a valuable testbed for developing and testing automation techniques.

%%%%%%%%%%%%%%%%%%%%%%%%%%%%%%%%%%%%%%%%%%%%%
%%%%%%%%%%%%%%%%%%%%%%%%%%%%%%%%%%%%%%%%%%%%%

While a variety of computational and machine learning (ML) techniques have been adopted for automation of QD device control, deep neural networks have proven especially capable of extracting useful information from CSDs~\cite{Zwolak20-AQD, Moon20-ATQ, Zwolak20-RBC, Czischek21-MNA, Ziegler23-AEC, Rao24-MAViS,  Marchand25-EAC, Moreno25-BML, Ziegler22-TAR}.
These advances have come amid a variety of progress in QD control, including bootstrapping and characterization~\cite{Kovach24-BATIS, Zubchenko24-ABQ, Ziegler22-TAR, Ziegler23-AEC}, state and charge tuning~\cite{Darulova19-ATQ, Zwolak20-AQD, Moon20-ATQ, Zwolak21-RBI, Ziegler22-TAR}, virtualization~\cite{Oakes2021-AVE, Rao24-MAViS, Lidiak25-VGD}, fine tuning for readout~\cite{Teske19-MFT, Carlsson25-ARF}, and gate operations~\cite{KatiraeeFar25-UEO} for QD qubit arrays of various sizes.

In this work, we present an automated protocol for extracting the electrostatic properties of coupled QD planar and bilayer systems directly from CSDs.
Our method combines deep neural networks with traditional image processing and fitting techniques to identify transition lines, track their motion across a series of CSDs, and compute their orientations and spacings.
From these features, we extract relative lever arms, charging voltages, and mutual energies, which we use to reconstruct the entire capacitive models of the underlying QD systems.
We benchmark the protocol using CSDs from a conventional, double-QD system in a planar germanium device.
We then apply the same workflow to a novel three-QD germanium bilayer device, demonstrating that we can automatically disentangle overlapping line families, infer interlayer mutual couplings, and quantify the limits of the constant-capacitance approximation in strong-coupling regimes.
In both cases, our protocol enables rapid, robust extraction of electrostatic parameters across large CSD datasets, providing a device-agnostic tool for characterizing and iterating QD architectures.

%%%%%%%%%%%%%%%%%%%%%%%%%%%%%%%%%%%%%%%%%%%%%
\begin{figure*}[t]
    \centering
    \includegraphics[width=\textwidth]{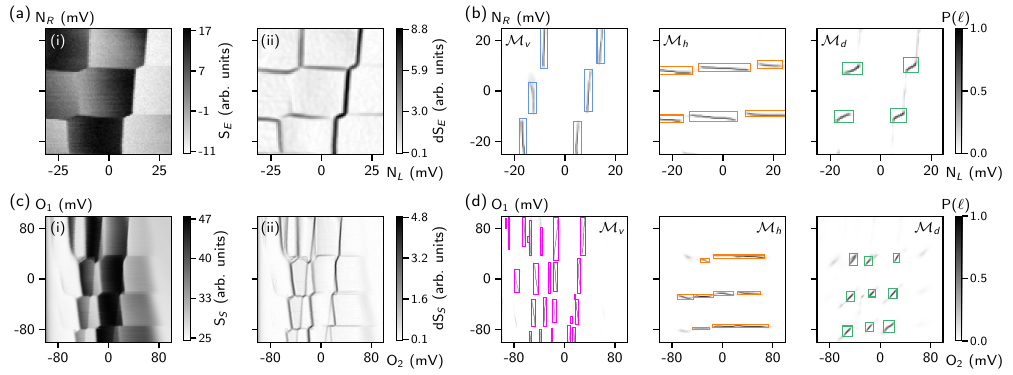}
    \caption{
    \textbf{The workflow of the ML-enabled CSD analysis.}
    A CSD typical of a (a-i) planar double-QD system and (c-i) a bilayer triple-QD system. 
    A CSD gradient, illustrated in (a-ii) for a planar device and in (c-ii) for a bilayer CSD, is passed through three U-Net pixel classifiers. 
    The outputs of each classifier are shown in (b) and (d), with high-intensity regions corresponding to pixels assigned a high probability of belonging to a given transition class.
    The final transitions are highlighted with the bounding boxes computed using a combination of thresholding and clustering.
    }
    \label{fig:models}
\end{figure*}
%%%%%%%%%%%%%%%%%%%%%%%%%%%%%%%%%%%%%%%%%%%%%

%%%%%%%%%%%%%%%%%%%%%%%%%%%%%%%%%%%%%%%%%%%%%%%%%%%%%%%%%%%%%
\section{Results}
\label{sec:results}
%%%%%%%%%%%%%%%%%%%%%%%%%%%%%%%%%%%%%%%%%%%%%%%%%%%%%%%%%%%%%
Our automated method for extracting physically relevant information from CSDs proceeds in three phases:
\begin{description}[topsep=3pt, itemsep=-3pt, leftmargin=15pt]
    \item[Phase 1:] \textit{ML-based transition detection and labeling.} 
    Identifying transition-line pixels in CSDs and assigning them to coarse transition classes using ML models and image processing.
    \item[Phase 2:] \textit{Geometric reconstruction of transition networks.} 
    Combining clustering, tracking, and other image processing tools to group segments into individual transitions, track them across a series of CSDs, and determine their angles and relative positions in gate-voltage space.
    \item[Phase 3:] \textit{Electrostatic parameter inference.} Mapping transition geometry to electrostatic parameters within a constant-capacitance model, including relative lever arms, charging energies, mutual couplings, and capacitance matrices.
\end{description}

Crucially, each phase operates only on abstracted representations of the CSDs (pixel-wise labels, line segments, their angles, and positions), rather than on device-specific details.
As a result, the same protocol can be applied across different device architectures and material stacks.
In what follows, we illustrate the whole workflow on two qualitatively different examples---a planar germanium double-QD system in a planar 2D array and a three-QD system in a germanium bilayer device---to demonstrate this device-agnostic character.

%%%%%%%%%%%%%%%%%%%%%%%%%%%%%%%%%%%%%%%%%%%%%%%%%%%%%%%%%%%%%
\subsection{ML-based transition detection and labeling} 
\label{ssec:ml_methods}
%%%%%%%%%%%%%%%%%%%%%%%%%%%%%%%%%%%%%%%%%%%%%
In the first phase of the protocol, we utilize U-Nets, a type of deep-learning model designed for image segmentation, to isolate and label individual transition lines in CSDs~\cite{Ronneberger15-UNET}; see the Methods section for details. 
We begin with CSDs measured as functions of two plunger gates, as shown in Fig.~\ref{fig:models}(a-i) and~\ref{fig:models}(c-i).
For the planar double-QD device, we follow the notation of Ref.~\cite{Rao24-MAViS} and define normalized plungers $\mathrm{N}_L$ and $\mathrm{N}_R$ corresponding to the effective left and right QD.
For the bilayer device, we define two orthogonalized plunger combinations $\mathrm{O}_1$ and $\mathrm{O}_2$, which primarily control the occupation of the QDs in the upper layer under the two physical plunger gates $\mathrm{P}_1$ and $\mathrm{P}_2$, respectively.

To enhance the visibility of transition lines, each CSD is converted to a gradient image;  see Fig.~\ref{fig:models}(a-ii) and~\ref{fig:models}(c-ii). 
These gradient images are then passed through three U-Net models~\cite{Ronneberger15-UNET}: $\mathcal{M}_\textit{v}$ for vertical transitions, $\mathcal{M}_\textit{h}$ for horizontal transitions, and $\mathcal{M}_\textit{d}$ for diagonal transitions.
The U-Nets are trained on either double-QD (for planar CSDs) or triple-QD (for bilayer CSDs) data simulated using QArray~\cite{vanStraaten24-QAR, vanStraaten24-QAC}, and they output a number between zero and one for each pixel, representing a probability that the pixel contains a transition of a given type.
The soft pixel-wise outputs are then converted to discrete transition segments by applying a fixed probability threshold, followed by connected-component analysis of the resulting binary masks; see the Supplementary Materials for details.
Each connected component is enclosed in a tight bounding box, which defines a candidate transition line segment.
Examples of these bounding boxes are overlaid on the classifier outputs in Fig.~\ref{fig:models}(b,d).

\begin{figure*}[t]
    \centering
    \includegraphics[width=\textwidth]{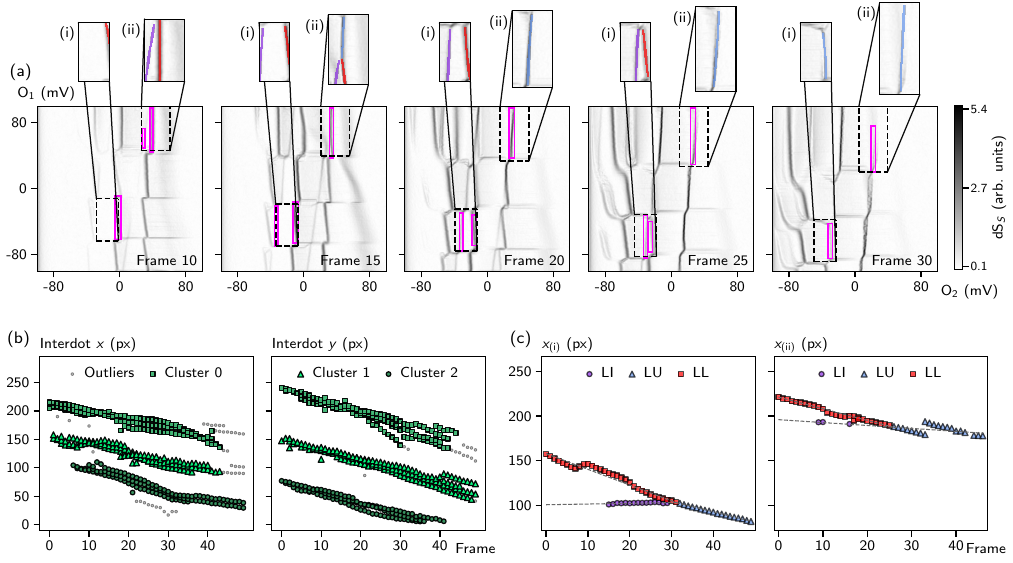}
    \caption{
    \textbf{Fine-grain classifications of the vertical lines in bilayer devices.}
    (a) A motion of two cells across five frames from a measurement series of bilayer CSDs. 
    Pink bounding boxes in each cell represent the transitions detected by the $\mathcal{M}_v$ model.
    Manually assigned LU (blue), LL (red), and LI (purple) labels for transition within the highlighted cells are shown in the insets. 
    (b) The clustered $x$- and $y$-coordinates of the interdot locations for the measurement series shown in (a). 
    (c) The change in the $x$-coordinates for transitions detected by the $\mathcal{M}_v$ model within the cells highlighted in (a). Colors indicate the fine-grain labels, while gray dashed lines indicate the 3-linear model fitted for each cell.
    }
    \label{fig:bilayer_cell_tracking}
\end{figure*}

For the planar CSDs, the classification of the transition maps directly onto their type, see Fig.~\ref{fig:models}(b). 
For bilayer CSDs, the vertical class includes left upper (LU) and left lower (LL) loading transitions, as well as the left interlayer transition (LI), all of which have similar sizes and orientations; see Fig.~\ref{fig:bilayer_cell_tracking}(a).
Distinguishing between these three types of transition requires an additional fine-grained categorization, as described in the next section.

%%%%%%%%%%%%%%%%%%%%%%%%%%%%%%%%%%%%%%%%%%%%%
%%%%%%%%%%%%%%%%%%%%%%%%%%%%%%%%%%%%%%%%%%%%%

%%%%%%%%%%%%%%%%%%%%%%%%%%%%%%%%%%%%%%%%%%%%%%%%%%%%%%%%%%%%%
\subsection{Geometric reconstruction of transition networks}
\label{ssec:geom_analysis}
%%%%%%%%%%%%%%%%%%%%%%%%%%%%%%%%%%%%%%%%%%%%%%%%%%%%%%%%%%%%%
For planar double-QD systems, the three U-Net-assigned classes map directly onto the underlying physical transition types: $\mathcal{M}_v$ identifies left-QD loading transitions, $\mathcal{M}_h$ identifies right-QD loading transitions, and interdot transitions are identified by $\mathcal{M}_d$.
In the bilayer CSDs, however, the transition lines detected by the $\mathcal{M}_v$ model---LU, LL, and LI---share similar sizes and orientations, necessitating an additional, fine-grained classification.
To accomplish this, we utilize the different lever arms between the gates and the QDs formed in the upper and lower quantum wells, as described below.

In Fig.~\ref{fig:bilayer_cell_tracking}(a), we show a series of bilayer CSDs, as the system is varied along a third axis in gate voltage space; in this case, we vary $\mathrm{vB}_{TL}$, a virtualized version of barrier gate $\mathrm{B}_{TL}$ in Fig.~\ref{fig:intro}(e).
Empirically, we observe that as the third axis is varied, two of these lines (LL and LI) approach and then merge into a third (LU) as a function of the sweep parameter, due to the different lever arms between this gate and the LU and LL dots.
To assign consistent labels to the LL, LU, and LI lines, it is therefore essential that we track the same set of transitions across consecutive frames, as illustrated in the insets of Fig.~\ref{fig:bilayer_cell_tracking}(a).
To assign labels in a device-agnostic way, we exploit the systematic way in which these lines merge and separate across a series of CSDs.

We achieve this by constructing a reference frame that divides each CSD into a grid of small, potentially overlapping rectangular \textit{cells}, using the locations of the interdot transitions (transitions from the LU or LL dots to the R dot) output by the model $\mathcal{M}_d$ for each frame in the series.
In the orthogonalized plunger space, the interdot transitions are expected to form an approximately orthogonal grid, which allows us to group interdot segments as a function of frame index independently for O$_1$ and O$_2$, see Fig.~\ref{fig:bilayer_cell_tracking}(b).
This yields robust tracklets for each interdot transition, even in the presence of occasional missed detections or spurious segments~\footnote{At this stage, we discard any interdot clusters with less than 20 members, to prevent spurious classifications.}.
By fitting the $x$- and $y$-coordinates of the interdot segments in each cluster as a function of frame index, we obtain smooth trajectories for each interdot line, which we then use to define and track the individual cells.

To automatically extract fine-grained labels, we map the $x$-coordinates for all vertical transitions within each cell series as a function of frame, as shown in Fig.~\ref{fig:bilayer_cell_tracking}(c).
We then fit the resulting data to a \textit{tri-linear} model---a piecewise-linear representation in which two lines intersect and subsequently follow a common branch, mimicking the empirically observed behavior in which the LL and LI transitions converge onto LU as the sweep parameter is varied.
The three linear branches of the tri-linear model correspond one-to-one with the three vertical transition types, providing a principled way to consistently relabel the left transitions as LU, LL, or LI across frames and devices using only geometric information.
Additional implementation details and edge cases are discussed in the Supplementary Material.

With transition labels in hand, we next extract the geometric properties of the CSD.
First, we extract the slopes of all transition lines by computing the orientation of all line segments in each bounding box.
Following Ref.~\cite{Rao24-MAViS}, we apply a Hough transform $H(\theta, d)$ to the pixels within each box and define a line-strength functional
\begin{equation} \label{eq:hsq}
    H_\mathrm{sq} (\theta) = \sum\nolimits_{d} H^2(\theta, d),
\end{equation}
which is sharply peaked at the dominant line orientation.
To extract an angle from this function, we rescale $\tilde H_\mathrm{sq}(\theta)$ to have unit integral, which we fit to a Cauchy probability density function over the range $0$ to $\pi$ for vertical class and $-\pi/2$ to $\pi/2$ for horizontal class~\footnote{While the particular form of this fit function is unimportant, we find that the strong peak and fat tails of a Cauchy distribution work well for transition line slopes.}:
\begin{equation}
    f_{\gamma, \theta_0}(\theta) = \left(\pi \gamma \left\{ 1 + \left[(\theta - \theta_0)/\gamma \right]^2 \right\}\right)^{-1}.
\end{equation}
The best-fit value of $\theta_0$ is the resulting angle of the transition line.

\begin{figure}[t]
    \centering
    \includegraphics[width=\columnwidth]{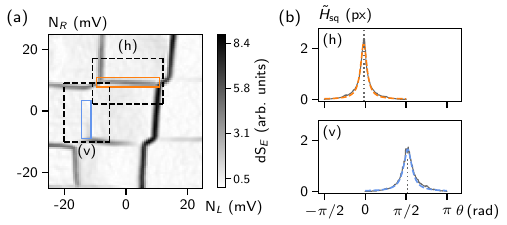}
    \caption{
    \textbf{Horizontal and vertical line transition in planar CSD.}
    (a) A right (h) and left (v) loading transition line for a sample planar double-QD CSD with colored bounding boxes, each encapsulated by the corresponding cell (black dashed boxes). 
    (b) Hough analysis of the transition line slopes highlighted in (a). 
    Gray lines indicate the function $\tilde H_\mathrm{sq}$, while dashed colored lines indicate the Cauchy fits. 
    Gray dotted lines indicate the best-fit angle $\theta_\mathrm{max}$. 
    }
    \label{fig:slope_cluster}
\end{figure}

Figure~\ref{fig:slope_cluster}(a) shows a planar CSD with two transition lines highlighted with bounding boxes colored based on the transition type.
The corresponding $\tilde H_\mathrm{sq}$ together with the resulting Cauchy fits are plotted in Fig.~\ref{fig:slope_cluster}(b). 
The best-fit angles $\theta_0$ are indicated with gray dotted lines.
Using the slopes of each transition line, we can extract the ratio of lever arms for each plunger gate to each dot in the system.

To extract remaining capacitive information, including charging and mutual energies, we need to understand how all transitions within a CSD relate to one another.
To do so, we utilize the interdot-based cells encapsulating charge transitions, shown as dashed black boxes in Fig.~\ref{fig:bilayer_cell_tracking}(a) for the bilayer CSDs and in Fig.~\ref{fig:slope_cluster}(a) for the planar CSDs.
All transition lines within a given cell are associated with the interdots that define that cell.
By mapping all cells in the plunger-plunger space and tracking them as the third axis is varied, we can extract the spatial relationships between transition lines across all CSDs, as we describe in the Supplementary Material.

This analysis gives us a purely geometric and device-agnostic characterization of each CSD in terms of (i) the distribution of angles for each transition type and (ii) the transitions network defining the relative positions of neighboring lines, both within a given type and between different types.
These quantities can now be mapped to electrostatic parameters.

%%%%%%%%%%%%%%%%%%%%%%%%%%%%%%%%%%%%%%%%%%%%%%%%%%%%%%%%%%%%%
\subsection{Electrostatic characterization of the quantum dot system}
\label{ssec:results_electrostatic}
%%%%%%%%%%%%%%%%%%%%%%%%%%%%%%%%%%%%%%%%%%%%%%%%%%%%%%%%%%%%%
Having obtained a device-agnostic geometric description of the CSDs, we now convert these quantities into electrostatic parameters using the constant-capacitance formalism described in the Methods.
The automatically obtained orientations and separations of lines in the transition networks are mapped to lever arms, as well as charging and mutual voltages (i.e., the spacings of transition lines in gate-voltage space).
These, in turn, are related to dot-dot and gate-dot capacitances.
To validate the automated characterization, we compare all results with those obtained from two manually labeled CSD series for each device. 
When discussing distributions, we use medians to reduce sensitivity to occasional misclassifications or outliers.

For the planar double-QD device, the line orientations and the transition networks provide sufficient information to extract the complete capacitive model underlying the coupled QD system.
For the bilayer triple-QD system, however, one of the charging voltages is not represented in the data, and therefore, we cannot fully reconstruct the capacitive model~\footnote{In our data, we do not observe successive LL lines, without an intermediate LI line. 
Thus, we cannot extract the left-lower charging energy. 
But, we note that, were such transitions visible in the data, our data-processing procedure would be able to extract this quantity.}.
Nonetheless, we can recover all three mutual energies in the system, including the nontrivial interlayer mutual energy.

%%%%%%%%%%%%%%%%%%%%%%%%%%%%%%%%%%%%%%%%%%%%%
%%%%%%%%%%%%%%%%%%%%%%%%%%%%%%%%%%%%%%%%%%%%%

%%%%%%%%%%%%%%%%%%%%%%%%%%%%%%%%%%%%%%%%%%%%%%%%%%%%%%%%%%%%%
\subsubsection{Transition angles and relative lever arms}
\label{sssec:angles_leverarms}
\begin{figure*}
    \centering
    \includegraphics[width=\textwidth]{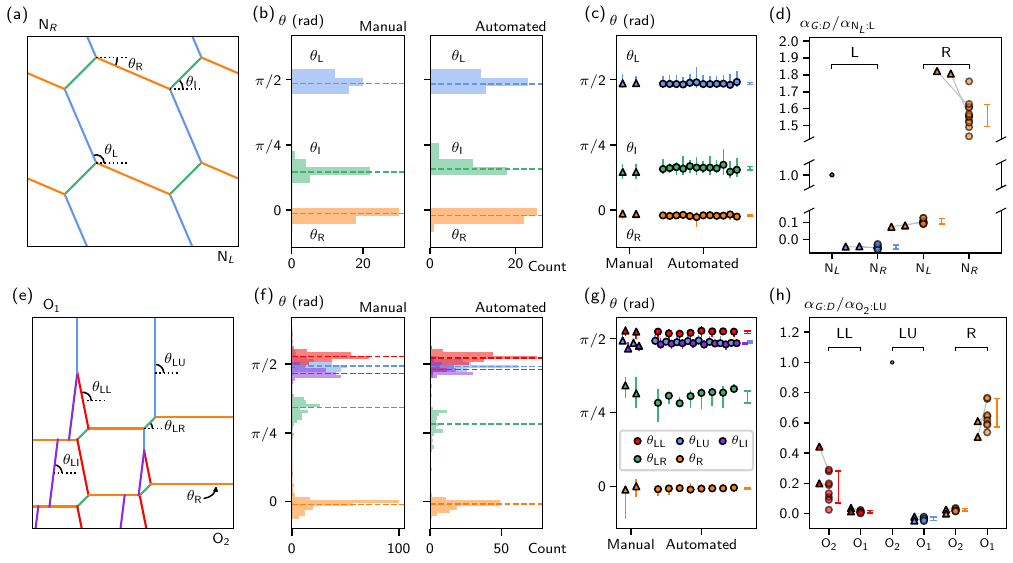}
    \caption{
    \textbf{Transition line slopes and relative lever arms for the planar and bilayer CSDs. }
    Schematic CSDs highlighting all relevant transition angles for an (a) planar and (e) bilayer CSD. 
    Distributions of transition angles for a sample CSD series for the (b) planar and (f) bilayer CSD.
    The two histograms compare angles extracted from manually labeled CSDs (left) with angles from automatically characterized CSDs (right).
    Dashed lines indicate median angles. 
    Transition angle distributions across the (c) 12 planar and (g) 8 bilayer CSD series.
    For each series, the points indicate medians, and the error bars represent the 10-90 percentile range. 
    The relative lever arms for each dot-gate pair in the (d) planar and (h) bilayer QD systems, computed from the median angles in (c) and (g), respectively.
    For the planar CSDs analysis, we fix $\alpha_{\mathrm{N}_L : \mathrm{L}}{=}1$, and for the bilayer CSDs analysis  $\alpha_{\mathrm{O}_2 : \mathrm{LU}}{=}1$, as indicated with small circles.
    The remaining lever arms are computed relative to these values. 
    In (c), (d), (g), and (h), circles (triangles) indicate results for automatically characterized (manually labeled) datasets; error bars on the right indicate the 10-90 percentile range of the medians of the automatically characterized CSD series.
    In (c) and (d), the two manually labeled datasets correspond to the first two automated datasets on the left.  
    In (d) and (h), gray lines connect the manually labeled CSD series with the corresponding automatically characterized series.
    }
    \label{fig:angles_combined}
\end{figure*}

%%%%%%%%%%%%%%%%%%%%%%%%%%%%%%%%%%%%%%%%%%%%%
Within the constant-capacitance model, each loading-line angle is determined by the ratio of lever arms between the two scanned gates and the corresponding QD.
We use the line orientations to compute relative lever arms for each pair of plunger and QD.
While we already have orthogonalized gates for our experimental CSDs, we do not rely on this orthogonalization, ensuring our method works more broadly.
Figure~\ref{fig:angles_combined} summarizes the aggregate distributions of all loading and interdot transition orientations, and the resulting lever arms, for all series of planar and bilayer CSDs.
The relevant transition angles within the CSDs for the planar devices are illustrated in Fig.~\ref{fig:angles_combined}(a) and for the bilayer devices in Fig.~\ref{fig:angles_combined}(e).

%%%%%%%%%%%%%%%%%%%%%%%%%%%%%%%%%%%%%%%%%%%%%
%%%%%%%%%%%%%%%%%%%%%%%%%%%%%%%%%%%%%%%%%%%%%

Figure~\ref{fig:angles_combined}(b) shows histograms of the three transition angles $\theta_\mathrm{L}$, $\theta_\mathrm{R}$, and $\theta_\mathrm{I}$, extracted from one series of 8 planar CSDs.
We note that the interdot transition angles $\theta_\mathrm{I}$ for planar CSDs are not extracted through the Hough transform of the detected inter-dot lines.
Instead, we leverage the transition networks and use the neighboring left- and right-loading transitions, along with their angles, to triangulate the orientation of the interdot lines (see Methods).
The automated characterization results, depicted on the right, show excellent agreement with those computed from manually labeled diagrams, shown on the left.
This strong agreement is further confirmed across all planar CSD series, with differences that are small compared to the intrinsic spread within each transition type; see Fig.~\ref{fig:angles_combined}(c).

In the bilayer device, the presence of additional lines complicates the picture.
For example, we must ensure that we only include interdot transitions between the intended pair of dots (e.g., LU and R) when computing angles such as $\theta_{\mathrm{LR}}$.
To avoid including transitions between the LL and R dots, we ensure each interdot has an LU loading line directly above and below it, as depicted in Fig.~\ref{fig:angles_combined}(e).
We achieve this by analyzing the positions of each line within the transition network, which enables us to identify the adjacent loading lines to each interdot.
In all cases, the association of loading lines to reference interdots is based purely on geometry and does not depend on device-specific modeling assumptions.
Also, to avoid distortions due to strong coupling between the LU and LL QDs, we only consider the two right-most LU, LL, LI, or LR transitions and the three rightmost R transitions within each bilayer CSD.
In Fig.~\ref{fig:angles_combined}(f), we compare histograms of transition angles for a manually-labeled (left) and a fully automatically characterized (right) series of 50 CSDs.
Despite additional transition lines complicating the classification procedure, we still see very good agreement between the manual and automated results, confirming the power of our procedure.
Distributions of transition angles for all series of bilayer CSDs are presented in Fig.~\ref{fig:angles_combined}(g).

%%%%%%%%%%%%%%%%%%%%%%%%%%%%%%%%%%%%%%%%%%%%%
\begin{figure*}[t]
\centering
  \begin{minipage}[c]{0.68\textwidth}
    \includegraphics{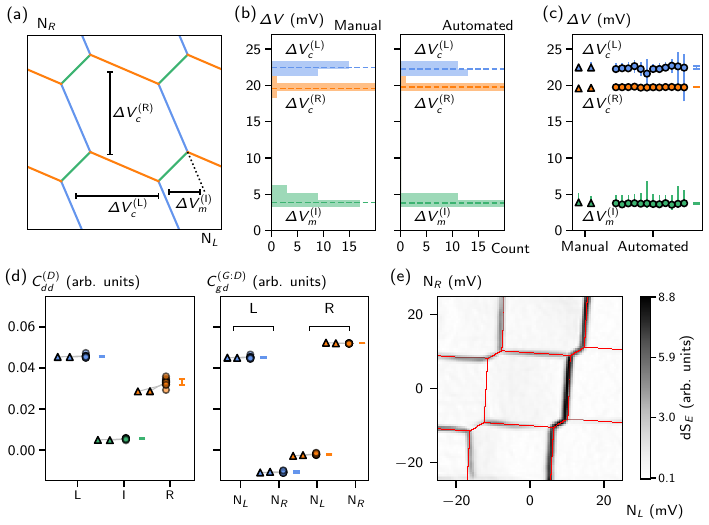}%
  \end{minipage}
  \hfill
  \begin{minipage}[c]{0.30\textwidth}
    \caption{
    \textbf{Analysis of charging and mutual voltages for planar QD device.} 
    (a) Schematic CSD with the charging voltages, $\Delta V_c^\mathrm{(L)}$ and $\Delta V_c^\mathrm{(R)}$, and mutual voltage, $\Delta V_m^\mathrm{(I)}$, labeled. 
    (b) Distributions of the charging and mutual voltages for a sample CSD series. 
    The two histograms compare angles extracted from manually labeled CSDs (left) with angles from automatically characterized CSDs (right).
    Dashed lines indicate medians.
    (c) Charging and mutual voltage distributions across 12 datasets.
    The markers indicate medians, and the error bars represent the 10-90 percentile range.
    (d) The dot-dot capacitances $C_{dd}$ (left) and the gate-dot capacitances $C_{gd}$ (right) computed using the median charging and mutual voltages in (c) and the median lever arms shown in Fig.~\ref{fig:angles_combined}(d). 
    (e) A CSD simulated using the extracted capacitance values (red lines) compared with experimental data (grayscale).
    }
    \label{fig:mut_charg_planar}
  \end{minipage}
\end{figure*}
%%%%%%%%%%%%%%%%%%%%%%%%%%%%%%%%%%%%%%%%%%%%%

By combining angles across different line families, we can now extract relative lever arms $\alpha_{G : D}$ from each plunger $G$ to each dot $D$ in the system (see Methods for details on this calculation). 
Since we can only determine relative lever arms, we must choose one lever arm to fix.
For consistency, we choose to fix the lever arm of the left plunger on the left QD for both devices, that is, $\alpha_{\mathrm{N}_L : \mathrm{L}} = 1$ for the planar double-QD data, and $\alpha_{\mathrm{O}_2 : \mathrm{LU}} =1$ for the triple-QD bilayer data, indicated with blue dots in Fig.~\ref{fig:angles_combined}(d) and (h).
This choice sets an energy scale for a device, from which we can compute the remaining lever arms.
Having made this choice, we extract the remaining relative lever arms using the median angles from each CSD series.
The resulting lever arms for planar and bilayer data are shown as colored circles in Fig.~\ref{fig:angles_combined}(d) and (h), respectively.
Error bars on the right indicate the 10-90 percentile range of the automated results.
Lever arms computed for the manually-labeled CSDs are shown as triangles, and gray lines indicate which automated result corresponds to each manual calculation.

For both planar and bilayer data, the results for manually labeled CSDs agree reasonably well with those from the fully automated procedure.
Moreover, the results make intuitive sense: each plunger is most strongly coupled to the QD immediately below it.
For the bilayer QD system, the plungers are more strongly coupled to the top-layer QD than to the bottom-layer QD, again in agreement with the intuition about this device.

For both systems, we observe some spread between the manual and automated estimates, most likely due to small biases in angle determination and slight variation across datasets.
Uncertainties in the right-QD lever arms are made larger by errors in the interdot angles, especially in the bilayer device.
In addition to the natural uncertainty inherent in experimental data, this spread partly reflects deviations from the constant-capacitance assumption: as gate voltages are varied, the underlying capacitive couplings do not need to remain fixed in experimental devices, while our modeling assumes constant couplings.
We analyze these deviations in more detail in the Discussion.
Nonetheless, our procedure provides an accurate and fully automated first-order estimate of the relevant electrostatic parameters.

%%%%%%%%%%%%%%%%%%%%%%%%%%%%%%%%%%%%%%%%%%%%%
\begin{figure*}
    \centering
    \includegraphics[width=\textwidth]{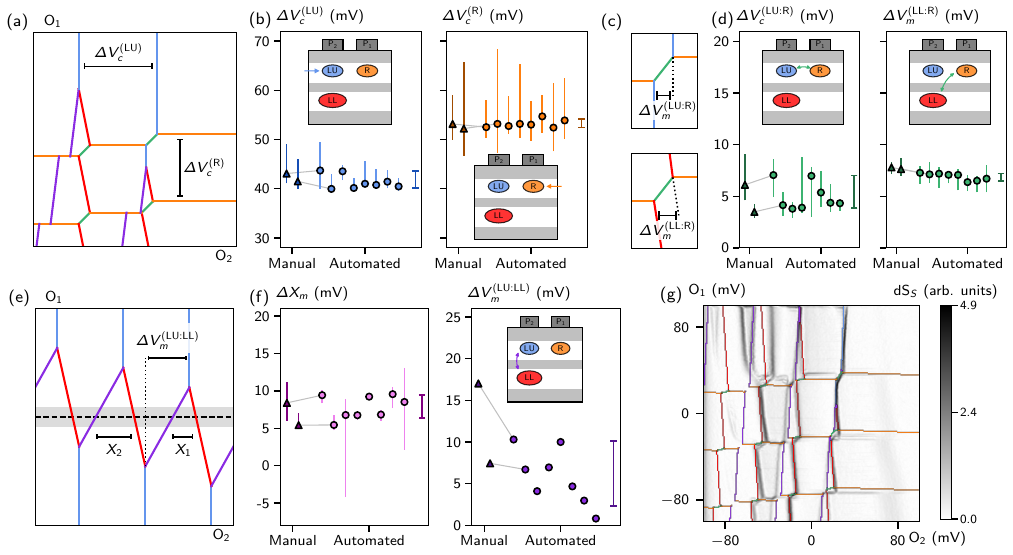}
    \caption{
    \textbf{Analysis of charging and mutual voltages for bilayer QD device.} 
    (a) Schematic CSD showing the $\Delta V_c^\mathrm{(LU)}$ and $\Delta V_c^\mathrm{(R)}$ charging voltages.
    (b) Distributions of $\Delta V_c^\mathrm{(LU)}$ (left) and $\Delta V_c^\mathrm{(R)}$ (right) across all bilayer CSD series.
    In all plots, the markers indicate the median, and the error bars indicate the 10-90 percentile range. 
    Error bars on the right indicate the 10-90 percentile range of the resulting automatically characterized CSD series medians.
    The gray lines connect the manually labeled CSD series with the corresponding automatically characterized series.
    (c) Schematic CSDs showing the two mutual voltages between left and right dots $\Delta V_m^\mathrm{(LU:R)}$ (top) and $\Delta V_m^\mathrm{(LL:R)}$ (bottom).
    (d) Distributions of $\Delta V_m^\mathrm{(LU:R)}$ (left) and $\Delta V_m^\mathrm{(LL:R)}$ (right) across all bilayer CSD series.
    (e) Schematic CSDs without any right-dot transitions, highlighting the spacing between LI and LL lines in adjacent cells ($X_1$ and $X_2$) used to compute the mutual voltage between upper and lower left QD $\Delta V_m^\mathrm{(LU:LL)}$. 
    (f) Distributions of $\Delta X_m$ (left) and the resulting $\Delta V_m^\mathrm{(LU:LL)}$ (right) across all bilayer CSD series.
    (g) A constant-capacitance CSD (colored lines), simulated using the global median-of-median automatically extracted lever arms and charging voltages, overlaid on an experimental CSD (grayscale).
    }
    \label{fig:mut_charg_bilayer}
\end{figure*}
%%%%%%%%%%%%%%%%%%%%%%%%%%%%%%%%%%%%%%%%%%%%%

%%%%%%%%%%%%%%%%%%%%%%%%%%%%%%%%%%%%%%%%%%%%%%%%%%%%%%%%%%%%%
\subsubsection{Charging and mutual voltages}
\label{sssec:charging}
%%%%%%%%%%%%%%%%%%%%%%%%%%%%%%%%%%%%%%%%%%%%%
Determining charging and mutual voltages for both planar and bilayer devices requires more complete models of the CSDs than those used for the lever arms.
To estimate charging voltages, we need to compute the spacing between adjacent transition lines along each axis in the CSD, shown in Fig.~\ref{fig:mut_charg_planar}(a) and Fig.~\ref{fig:mut_charg_bilayer}(a) for a planar and bilayer system, respectively. 
Mutual voltages are defined as the lateral offsets between loading lines across an interdot transition, and are related to the mutual capacitance between QDs; see Fig.~\ref{fig:mut_charg_planar}(a) for a planar system and Fig.~\ref{fig:mut_charg_bilayer}(c) and (e) for a bilayer system.
To determine which lines to evaluate, we utilize a comprehensive spatial map of the CSD, i.e., the complete transition network introduced earlier.
     
Because the transition lines are not, in general, aligned with either gate axis, we compute these spacings by projecting the vector between two points on the lines along the appropriate direction.
To estimate the horizontal spacing between two transition lines with centers $(x_1, y_1)$ and $(x_2, y_2)$, and angles $\theta_1$ and $\theta_2$, we use the following equation:
\begin{equation} \label{eq:horiz_dist}
    \Delta V_\mathrm{L} = -\frac{[y_2 - y_1]}{\tan\langle\theta\rangle} + [x_2 - x_1],
\end{equation}
where $\langle\theta\rangle = (\theta_1 + \theta_2) / 2$.
An analogous expression is used for vertical spacings~\cite{Rao24-MAViS},
\begin{equation} \label{eq:vert_dist}
    \Delta V_\mathrm{R} = -[x_2 - x_1]\tan \langle \theta \rangle + [y_2 - y_1]
\end{equation}
with the roles of the two axes interchanged.
Under the constant-capacitance assumption, all transitions of a given family are expected to have the same angle, so that $\theta_1 \approx \theta_2$.

Applying this procedure to a sample planar CSD series used in Fig.~\ref{fig:angles_combined}(a–d) yields distributions of the left- and right-dot charging voltages, $\Delta V_c^\mathrm{(L)}$ and $\Delta V_c^\mathrm{(R)}$, and the mutual voltage, $\Delta V_m^\mathrm{(I)}$ shown in Fig.~\ref{fig:mut_charg_planar}(b).
Once again, the histograms for automated characterization (right) show excellent agreement with those computed from manually labeled diagrams (left).
This strong agreement persists when we aggregate results across all planar CSD series, again with differences relatively small compared to the intrinsic spread within each voltage type; see Fig.~\ref{fig:mut_charg_planar}(c).

Using the resulting charging voltages, Fig.~\ref{fig:mut_charg_planar}(c), together with the previously extracted lever arms, Fig.~\ref{fig:angles_combined}(d), we can calculate a complete electrostatic characterization of the planar device, including the dot–dot capacitance matrix $C_{dd}$ (defined up to an overall scale) and the gate–dot capacitance matrix $C_{gd}$.
The diagonal elements of $C_{dd}$, $C_{dd}^\mathrm{(L)}$ and $C_{dd}^\mathrm{(R)}$, describe the total capacitance of the left and right QD, respectively, while the off-diagonal element $C_{dd}^\mathrm{(I)}$ encodes their mutual capacitive coupling.
As described in the Methods, the inverse matrix $C_{dd}^{-1}$ is populated by the charging and mutual energies, which are related to the measured charging and mutual voltages, $\Delta V_c$ and $\Delta V_m$, through the corresponding lever arms.
Elements of $C_{dd}$ for the planar device, extracted for each series of CSDs, are shown in Fig.~\ref{fig:mut_charg_planar}(d) for both manually labeled (triangles) and automated (circles) datasets.

Using the same lever arms and $C_{dd}$, we extract the elements of matrix $C_{gd}$, where $C_{gd}^{(G:D)}$ describes the capacitive coupling between gate $G$ and dot $D$.
Results for each series of planar CSDs are shown in Fig.~\ref{fig:mut_charg_planar}(d).
These results are physically intuitive---each plunger couples strongly to the dot beneath it and only weakly to the other dot---and the close agreement between manual and automated analyses indicates that our method reliably recovers these capacitances.

To validate the extracted capacitive model, we simulate a CSD using parameters obtained from our analysis.
Taking the median-of-medians lever arms and capacitances from Fig.~\ref{fig:angles_combined} and Fig.~\ref{fig:mut_charg_planar}, we simulate a CSD with QArray~\cite{vanStraaten24-QAR, vanStraaten24-QAC}.
The resulting CSD, superimposed on an experimental CSD, uses only global offsets in $x$ and $y$ as fitting parameters.
Although the experimental CSD exhibits some deviations from strict constant-capacitance behavior, the overall agreement between simulated and measured CSDs is excellent, providing strong evidence that our automated protocol accurately recovers the planar double-QD capacitive couplings.

Next, we analyze the performance of the automated characterization protocol on the bilayer device.
Extracting charging voltages follows the same logic as for planar systems, with Eq.~(\ref{eq:horiz_dist}) used to determine the left-upper QD charging voltage, $\Delta V_c^\mathrm{(LU)}$ and Eq.~(\ref{eq:vert_dist}) used to determine the right-QD charging voltage $\Delta V_c^\mathrm{(R)}$.
Figure~\ref{fig:mut_charg_bilayer}(b) shows the distributions of $\Delta V_c^\mathrm{(LU)}$ (left) and $\Delta V_c^\mathrm{(R)}$ (right) across all manually labeled (triangles) and automated characterized (circles) CSD series, with very good consistency.
Again, the individual error bars indicate the 10-90 percentile range for each CSD series, while the error bars on the far right indicate the 10-90 percentile range for the medians of the automatically characterized CSD series.
Since throughout the experimental data acquisition, we were not able to capture successive LL loading without an intermediate LI interlayer transition, we cannot extract the $\Delta V_c^\mathrm{(LU)}$, which means we are not able to reconstruct the complete $C_{dd}$ and $C_{gd}$ matrices for this system.
Nevertheless, the available charging voltages, combined with the mutual voltages, suffice to provide insight into certain interesting properties of the bilayer system, which we discuss in the next section.

%%%%%%%%%%%%%%%%%%%%%%%%%%%%%%%%%%%%%%%%%%%%%
\begin{figure*}
    \centering
    \includegraphics[width=\textwidth]{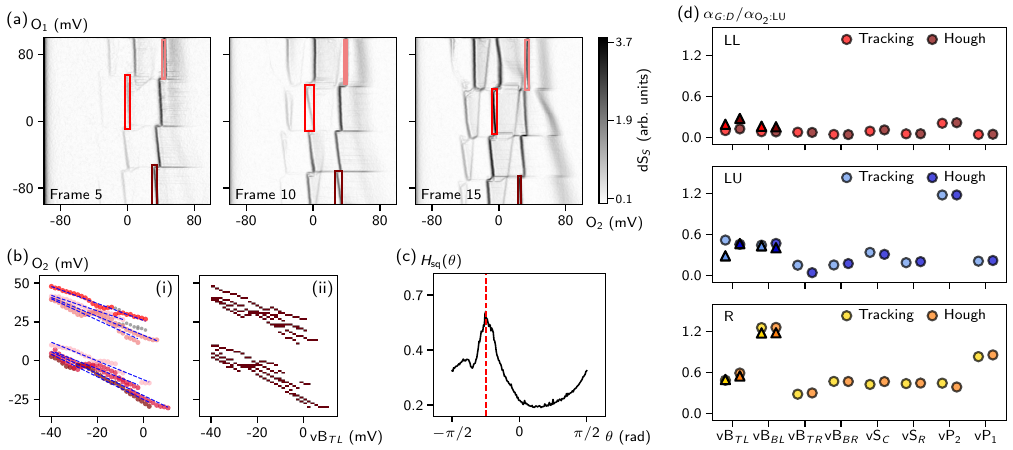}
    \caption{
    \textbf{Estimates for lever arms for each additional gate to each dot.} 
    (a) Three frames from a series of 50 CSD, where $\mathrm{vB}_{TL}$ at the center of each CSD is varied from \SI{-40}{\milli\volt} to \SI{40}{\milli\volt}. 
    The colors of bounding boxes represent the track across time for three selected transition lines. 
    (b) The $x$ ($\mathrm{O}_2$) coordinate of all transition lines, colored by the transition track,
    as a function of $\mathrm{vB}_{TL}$. 
    Panel (i) shows complete track data, with tracks discarded from analysis shown with small gray dots, while panel (ii) shows a binarized version of the dataset. 
    (c) $H_{sq}(\theta)$ computed for the binarized dataset, with  $\theta_\mathrm{max}$ indicated with a red dashed line. 
    (d) The lever arms computed for the LL (top), LU (middle), and R (bottom) QD. 
    Lighter markers indicate lever arms computed with the tracking method, and darker colors indicate lever arms computed with the Hough method. 
    Triangles indicate results on hand-labeled datasets.
    }
    \label{fig:leverarm_tracking}
\end{figure*}
%%%%%%%%%%%%%%%%%%%%%%%%%%%%%%%%%%%%%%%%%%%%%

There are three mutual voltages in the bilayer triple-QD system: two between the left and right QD and one between the top and bottom left QD.
We first analyze the mutual voltages between the left and right QD, illustrated schematically in Fig.~\ref{fig:mut_charg_bilayer}(c): $\Delta V_m^\mathrm{(LU:R)}$ (top) and $\Delta V_m^\mathrm{(LL:R)}$ (bottom).
We obtain these by computing the horizontal spacing between LU or LL loading lines separated across one interdot transition, using Eq.~(\ref{eq:horiz_dist}).
The resulting distributions, shown in Fig.~\ref{fig:mut_charg_bilayer}(d) for both manually labeled CSD (triangles) and fully automatically characterized CSDs (right), are much smaller than the corresponding charging voltages, as expected for relatively weak left–right coupling, and show good agreement between the two procedures.

We next estimate the mutual voltage between the LU and LL, $\Delta V_m^\mathrm{(LU:LL)}$, which cannot be read directly from the measured CSDs.
In an idealized CSD containing only LU, LL, and LI transitions, Fig.~\ref{fig:mut_charg_bilayer}(d), $\Delta V_m^\mathrm{(LU:LL)}$ would correspond to the horizontal spacing between LU lines across one LI line.
In practice, the finite measurement range and the presence of right-QD transitions mean we only observe narrow strips of this hypothetical diagram, highlighted in gray in Fig.~\ref{fig:mut_charg_bilayer}(e), confined between successive right-QD loading lines.
Using geometric relations implied by the constant-capacitance model, we thus instead estimate an intermediate quantity $\Delta X_m = X_2 - X_1$, where $X_1$ and $X_2$ are the horizontal spacings between LI and LL transitions, evaluated at the same $\mathrm{O}_1$ in adjacent cells, see Fig.~\ref{fig:mut_charg_bilayer}(e).
Details of this construction are given in the Methods.

Distributions of $\Delta X_m$ for manually labeled and automatically characterized series of CSDs are shown in the left panel in Fig.~\ref{fig:mut_charg_bilayer}(f), again showing good agreement.
Combining the median $\Delta X_m$ for each series with the median charging voltages and transition angles, we obtain the interlayer mutual voltage $\Delta V_m^\mathrm{(LU:LL)}$, with results summarized in the right panel in Fig.~\ref{fig:mut_charg_bilayer}(f).
Despite the complexity of this indirect estimate, the manual and automated values agree reasonably well, within the spread between different labeled datasets.
If the constant-capacitance approximation were valid, all labeled datasets would yield nearly identical results.
The spread within the labeled datasets suggests that deviations from constant capacitance, rather than the automation method, are a dominant source of variation.
%Despite the complexity of this indirect estimate, the manual and automated values agree within the spread between different labeled datasets, suggesting that deviations from the constant-capacitance approximation, rather than the automation itself, are the dominant source of variation.

Finally, to validate the automatically extracted capacitive parameters, we compare them to the experimental data via simulation.
Using the global median-of-medians capacitances and lever arms from all automated bilayer CSD series, we simulate a CSD and superimpose it on an experimental diagram, as shown in Fig.~\ref{fig:mut_charg_bilayer}(g).
For this simulation, we include coupling to a third gate in addition to $\mathrm{O}_1$ and $\mathrm{O}_2$, using the lever arms extracted for $\mathrm{vB}_{TL}$ on each QD (see the next section), and treat constant offsets along $\mathrm{O}_1$, $\mathrm{O}_2$, and $\mathrm{vB}_{TL}$ as fitting parameters.
Moreover, since we cannot extract the lower-QD charging energy $\Delta E_c^\mathrm{(LL)}$ from the data, in the simulation we set $\Delta E_c^\mathrm{(LL)} = 0.5\,\Delta E_c^\mathrm{(LU)}$ (see Supplementary Material for further details).
Nevertheless, the fit is excellent on the right-hand side of the diagram, indicating that the extracted parameters accurately capture the essential electrostatics of the bilayer device.

%%%%%%%%%%%%%%%%%%%%%%%%%%%%%%%%%%%%%%%%%%%%%%%%%%%%%%%%%%%%%
\subsection{Extended applications: Device-level characterization by tracking transition lines}
\label{ssec:tracking}
%%%%%%%%%%%%%%%%%%%%%%%%%%%%%%%%%%%%%%%%%%%%%%%%%%%%%%%%%%%%%
Thus far, we have utilized the automated workflow to characterize the electrostatics of coupled QD systems based on individual CSDs, with a focus on the lever arms relative to the CSD axes.
However, with CSDs acquired in a series as a function of auxiliary gate voltages, the same tools can be used to extract additional device-level information. 
By tracking the positions of transition lines across these series, we can directly measure lever arms between the QD and the auxiliary gates, providing a richer characterization of the device.
We then fit the QD-occupancy transition conditions to linear functions of the auxiliary gate voltages, extracting effective lever arms for all QD-auxiliary gate combinations.

Since all tracking and fitting procedures are device-agnostic, these extended applications do not require any device-specific modifications to the workflow.
Instead, they demonstrate that once transition lines are reliably detected, labeled, and tracked, a wide range of device- and material-level questions---such as mapping gate–dot coupling patterns or probing the influence of screening gates---can be addressed systematically by analyzing how the CSD evolves under controlled changes of external parameters.

By tracking the motion of loading lines relative to the horizontal (or vertical) axis of these CSD, we can extract lever arms for each QD, relative to a third gate axis.
The lever arm of a third gate potential $V_3$ on dot $D$ is computed using
\begin{equation}
    \alpha_{V_3 : D} = -\alpha_{G: D}\,v_G^{(D)}(V_3)
\end{equation}
where $V_G^{(D)}$ is the position of a loading line for dot $D$ along axis $G$, and $v_G^{(D)}(V_3)=dV_G^{(D)} / dV_3$ is the velocity of this loading line relative to $V_3$, and $G$ refers to either of the CSD axes, $\mathrm{O}_2$ or $\mathrm{O}_1$.

Since LU and LL transitions are nearly vertical, their motion along $\mathrm{O}_2$ is well captured by tracking their $x$-coordinates through the series, while for the horizontal R transitions, we track their $y$-coordinates along $\mathrm{O}_1$.
For LU, $\alpha_{\mathrm{O}_2:\mathrm{LU}}{=}1$ by definition.
For LL and R, we use the median-of-median lever arms $\alpha_{\mathrm{O}_2:\mathrm{LL}}$ and $\alpha_{\mathrm{O}_1:\mathrm{R}}$ across all automatically characterized CSD series reported in Fig.~\ref{fig:angles_combined}(h); for the hand-labeled data we use the average of the two labeled series.
The remaining challenge is to estimate the $v_G^{(D)}(V_3)$.
To achieve this, we employ two complementary methods to extract this motion: the Hungarian object-tracking algorithm~\cite{Kuhn55-HUN} and the Hough-transform-based approach, both of which are discussed in detail in the Methods section.

For the bilayer triple-QD device, we must track three sets of loading lines: LU, LL, and R.
Figure~\ref{fig:leverarm_tracking}(a) shows a series of CSDs with several LL transition lines highlighted (red bounding boxes) as a function of virtual barrier gate $\mathrm{vB}_{\mathrm{TL}}$, a virtualized version of the barrier gate $\mathrm{B}_{TL}$ shown in Fig.~\ref{fig:intro}(c).
The colors of these boxes indicate a track; we see that the Hungarian algorithm assigns the same transition to the same track across the CSDs.

Figure~\ref{fig:leverarm_tracking}(b) shows the position of all detected transitions along $\mathrm{O}_2$ versus $\mathrm{vB}_{TL}$ (i), with colored lines indicating accepted tracks (determined by the Hungarian algorithm), gray points indicating discarded ones, and the blue dashed lines represent the linear fits made for each track.
The binarized version of the $\mathrm{vB}_{TL}$ versus $\mathrm{O}_2$ plot is shown in Fig.~\ref{fig:leverarm_tracking}(b-ii) and the corresponding Hough transform in Fig.~\ref{fig:leverarm_tracking}(c).

Figure~\ref{fig:leverarm_tracking}(d) summarizes the resulting lever arms for LL (top), LU (middle), and R (bottom) across all eight datasets, with all values defined relative to $\alpha_{\mathrm{O}_2:\mathrm{LU}} = 1$.
The tracking- and Hough-based estimates (light and dark circles, respectively) are consistent with each other, and both agree with values obtained from hand-labeled data (triangles), lending further confidence to the automated analysis.

The resulting lever arms are also physically reasonable.
Of the two virtual (but not orthogonal) plungers $\mathrm{vP}_2$ and $\mathrm{vP}_1$, the left-side dots (LU and LL) couple most strongly to $\mathrm{vP}_2$, while the right dot (R) couples most strongly to $\mathrm{vP}_1$, as expected.
Similarly, LU couples more strongly to the nearby barrier gates $\mathrm{vB}_{TL}$ and $\mathrm{vB}_{BL}$ than to $\mathrm{vB}_{TR}$ and $\mathrm{vB}_{BR}$, and R couples more strongly to barrier and screening gates ($\mathrm{vS}_C$, $\mathrm{vS}_R$), which are closer to the right plunger $\mathrm{O}_1$ [Fig.~\ref{fig:intro}(e)].
Finally, the upper-layer dots (LU and R) are generally more strongly coupled to the barrier and screening gates than the lower-layer dot (LL).
Taken together, these trends match our geometric intuition for the device layout and demonstrate that the measured lever arms are both reasonable and reproducible.

In our analysis, we focused on the more complex bilayer device, which is a good test case. 
Nonetheless, these methods apply equally well to the planar double-QD data.

%%%%%%%%%%%%%%%%%%%%%%%%%%%%%%%%%%%%%%%%%%%%%%%%%%%%%%%%%%%%%
\section{Discussion}
\label{sec:discussion}
%%%%%%%%%%%%%%%%%%%%%%%%%%%%%%%%%%%%%%%%%%%%%%%%%%%%%%%%%%%%%
In this work, we have developed a fully automated protocol to extract electrostatic properties of QD devices from their charge stability diagrams.
By combining ML-based pixel classification, geometric clustering, and object tracking, our workflow assigns labels to individual transition lines, determines their angles and positions, and uses this information to compute lever arms, charging and mutual voltages, and capacitance matrices.
By tracking the motion of transitions as a third gate is varied, the same tools further yield lever arms to additional barrier and screening gates.
We have demonstrated this framework on two qualitatively different systems: a planar double-QD and a more complex bilayer triple-QD.

%%%%%%%%%%%%%%%%%%%%%%%%%%%%%%%%%%%%%%%%%%%%%
\begin{figure}
    \centering
    \includegraphics[width=0.5\textwidth]{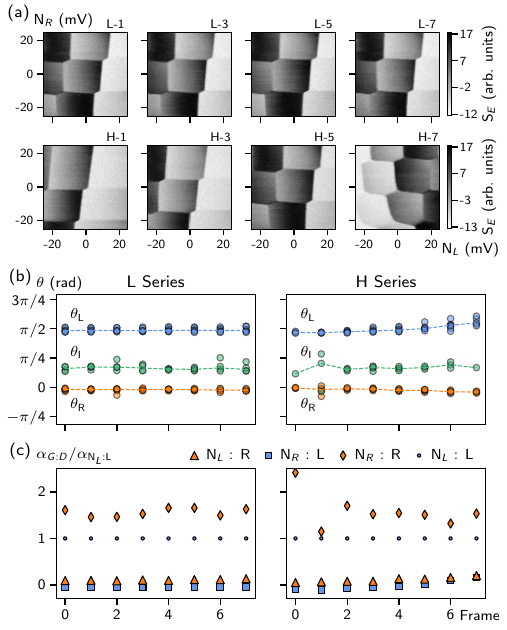}
    \caption{
    \textbf{Transition angles and lever arms over time.} 
    (a)~A series of CSD acquired within a ``low-coupling'' (L; top)  and ``high-coupling'' (H; bottom) regimes.
    (b)~Transition angles from the L (left) and H (right) CSD series.
    Dashed lines indicate the median values of each frame. 
    (c)~Lever arms computed from the median angles in (b) for the L (left) and R (right) CSD series.
    }
    \label{fig:mavis_changing_lever_arms}
\end{figure}
%%%%%%%%%%%%%%%%%%%%%%%%%%%%%%%%%%%%%%%%%%%%%

A central assumption in our analysis is that the device is well described by a constant-capacitance model, so that capacitive couplings remain fixed within and across CSD in a series.
This assumption enables us to aggregate information from multiple CSDs to obtain robust estimates of electrostatic parameters.
However, it is not universally valid.
To probe its limits, we examine two series of CSDs measured on the planar QD device, labeled in in Fig.~\ref{fig:mavis_changing_lever_arms}(a) with ``L'' (low coupling) and ``H'' (high coupling).

In the L series (top row), the CSDs remain relatively unchanged, and the corresponding transition angles and lever arms shown in Figs.~\ref{fig:mavis_changing_lever_arms}(c) and (d), respectively, remain essentially constant over time, consistent with a stable constant capacitance model description.
In contrast, the H series exhibits a clear evolution of the CSD as the device crosses over from weak to strong interdot coupling, visibly deforming the charge cells; see the bottom row in Fig.~\ref{fig:mavis_changing_lever_arms}(a).
This evolution is reflected in systematic trends in both angles and lever arms [see right panels in Figs.~\ref{fig:mavis_changing_lever_arms}(c,d)], including a pronounced increase in the cross lever arms $\alpha_{\mathrm{N}_{L}:\mathrm{R}}$ and $\alpha_{\mathrm{N}_{R}:\mathrm{L}}$, as expected when QD become more strongly coupled.

Since capacitive couplings vary across the high-coupling series, we do not expect a single constant capacitance model to fit all frames equally well.
Indeed, the corresponding automatically characterized CSD series (the sixth series in Figs.~\ref{fig:angles_combined}(c) and \ref{fig:mut_charg_planar}(c)) exhibits noticeably larger error bars than the other series.
This highlights an important practical point: meaningful electrostatic characterizations require that the coupling environment remain approximately unchanged across the data being aggregated.

At the same time, the analysis in Fig.~\ref{fig:mavis_changing_lever_arms} illustrates a key strength of our workflow.
Because the pipeline stores line-by-line geometry and tracks transitions across frames, it can be used not only to extract average parameters but also to monitor their time dependence, automatically flagging regimes in which the constant capacitance approximation begins to break down or identifying the crossover between weak- and strong-coupling regimes.

Our automated characterization workflow is readily extendable to different device layouts.
For planar arrays, analyzing successive pairs of QD yields a pairwise electrostatic characterization across the device.
While the triple-QD bilayer CSD required additional fitting to disentangle LU, LL, and LI transitions, this was also achieved automatically by exploiting their characteristic motion across a series of CSDs.

The particular geometry of the bilayer device measured in this work is not essential; the same strategy can be generalized to other triple-QD systems controlled primarily by two plunger gates.
This is particularly relevant for larger-scale architectures, such as the proposed crossbar-style arrays in which multiple QDs share common control gates~\cite{Veldhorst17-SCA, Li18-CNQ, Borsoi22-QCA}, for which CSDs will naturally contain transitions from more than two QDs.
In such settings, we expect our automated CSD analysis to provide a functional building block for systematic electrostatic characterization, both as a standalone diagnostic and as a component within larger autotuning and control frameworks.

%%%%%%%%%%%%%%%%%%%%%%%%%%%%%%%%%%%%%%%%%%%%%%%%%%%%%%%%%%%%%
\section{Methods}
\label{sec:methods}
%%%%%%%%%%%%%%%%%%%%%%%%%%%%%%%%%%%%%%%%%%%%%%%%%%%%%%%%%%%%%
In this section, we detail the procedures used to obtain, process, and analyze the CSDs considered in this work.
We first describe the bilayer device used in the experiments and the experimental setup, followed by details of the U-Net pixel classifiers and the preprocessing steps used to detect and label transition-line pixels in CSDs.
We then summarize the constant-capacitance formalism and show how transition angles and spacings are mapped to lever arms, charging and mutual voltages, and gate–dot capacitances.
Subsequent subsections provide implementation details for key geometric constructions, including the extraction of interdot angles in double-QD CSDs, the computation of interlayer mutual voltages in the bilayer device, and the tracking of transition-line motion across a series of CSDs used to infer lever arms to additional gates.

%%%%%%%%%%%%%%%%%%%%%%%%%%%%%%%%%%%%%%%%%%%%%
\begin{figure}
    \centering
    \includegraphics[width=8cm]{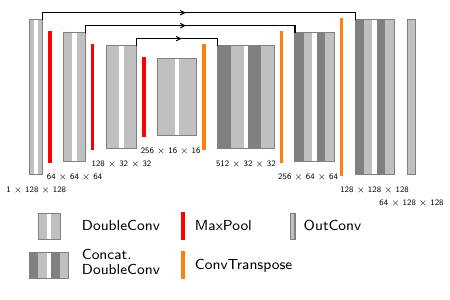}
    \caption{
    \textbf{Schematic illustration of the U-Nets.}
    In the downsampling leg of the network, the input image is passed through a succession of \texttt{DoubleConv} layers.
    In the upsampling branch, each convolution is followed by a 2D \texttt{ConvTranspose}.
    The outputs of the downsampling branch are additionally concatenated with the convolutional layers in the upsampling branch, as indicated by black arrows. 
    A sigmoid activation follows the final convolution layer. 
    The dimensionality of the input to each convolutional layer (channels~$\times$~width~$\times$~height) is indicated below each layer.
    }
    \label{fig:unet}
\end{figure}
%%%%%%%%%%%%%%%%%%%%%%%%%%%%%%%%%%%%%%%%%%%%%

%%%%%%%%%%%%%%%%%%%%%%%%%%%%%%%%%%%%%%%%%%%%%%%%%%%%%%%%%%%%%
\subsection{Device fabrication and experimental setup}
\label{ssec:exp}
%%%%%%%%%%%%%%%%%%%%%%%%%%%%%%%%%%%%%%%%%%%%%
The bilayer device measured in this work is fabricated on a Si$_{0.2}$Ge$_{0.8}$(/Ge/Si$_{0.2}$Ge$_{0.8}$)$_2$ heterostructure. 
The double quantum well (DQW) is grown on a silicon virtual substrate. 
The upper germanium QW has a nominal thickness of 10~\si{\nano\meter}, while the lower QW has a thickness of 16~\si{\nano\meter}; the two wells are separated by a 10~\si{\nano\meter} SiGe spacer.
Above the DQW, a 55~\si{\nano\meter} SiGe spacer and a sacrificial Si cap layer separate the heterostructure from the metallic gate stack.

Fabrication of the metallic gates begins with the ohmic contacts, which are defined using electron-beam lithography (EBL) and formed from 30~\si{\nano\meter} Pt.
During a subsequent thermal anneal at $400^\circ$C, the Pt diffuses down to the upper QW to form ohmic contacts.
The remaining gate layers shown in Fig.~\ref{fig:intro}(e) are then defined by repeating cycles of EBL, metal deposition, and oxide growth.
Together, these gates electrostatically define QD in the DQW.
More details about the device and the fabrication process can be found in Ref.~\cite{Tidjani25-3DA}.

All measurements are performed in a Bluefors LD400 dilution refrigerator with a base temperature of $T_{\text{base}} = 10$~\si{\milli\kelvin}.
DC voltages for each metallic gate are supplied by home-built serial peripheral interface (SPI) DAC modules. 
Radio-frequency (RF) pulses generated by a Qblox Cluster arbitrary waveform generator (AWG) are combined with the DC bias using on-PCB bias tees. 
The ohmic contacts of the charge sensors are connected to a resonant tank circuit that incorporates a custom-fabricated niobium–titanium–nitride (NbTiN) inductor for RF reflectometry.

%%%%%%%%%%%%%%%%%%%%%%%%%%%%%%%%%%%%%%%%%%%%%%%%%%%%%%%%%%%%%
\subsection{Machine learning models}
\label{ssec:unet}
%%%%%%%%%%%%%%%%%%%%%%%%%%%%%%%%%%%%%%%%%%%%%
The ML models employed in this work are the U-Net pixel classifiers, schematically illustrated in Fig.~\ref{fig:unet}.
The networks have an autoencoder-like structure, with a downsampling (encoder) branch followed by an upsampling (decoder) branch, and skip connections between corresponding layers.

The encoder contains three \texttt{DoubleConv} blocks, each followed by a 2D \texttt{MaxPool} layer with kernel size 2, which halves the spatial dimensions and doubles the number of channels.
Each \texttt{DoubleConv} block consists of a 2D convolution with kernel size three and padding one, followed by a ReLU activation, repeated twice.
The bottleneck consists of a fourth \texttt{DoubleConv} block followed by a \text{ConvTranspose} layer (kernel size 2, stride 2) that upsamples by a factor of two and halves the number of channels.

The decoder contains two additional upsampling steps, with each \texttt{ConvTranspose} layer followed by concatenation with the corresponding encoder feature map skip connection (indicated by black arrows in Fig.~\ref{fig:unet}) and a \texttt{DoubleConv} block.
The final \texttt{DoubleConv} output is passed through a 1$\times$1 convolution to produce a single-channel image with the original spatial dimensions, and a sigmoid activation mapping it to $[0,1]$.

The U-Nets are trained on $128{\times}128$ pixel CSD gradient images simulated with QArray~\cite{vanStraaten24-QAR, vanStraaten24-QAC}.
Planar double-QD CSDs from Ref.~\cite{Rao24-MAViS} are originally $81{\times}81$ pixels; these are upsampled to $128 \times 128$ before entering the networks.
Gradients are computed using \texttt{scipy.ndimage.gaussian\_gradient\_magnitude}~\cite{Virtanen20-SCI} with \texttt{sigma}~${=1}$, and the resulting images are normalized to the range $[0,1]$.

Bilayer triple-QD CSDs are originally $300 \times 300$ pixels.
For the network that detects left-class transitions, we directly extract $128{\times}128$ crops.
For the networks that detect right-and diagonal-class transitions, we take $256{\times}256$ crops and resize them to $128{\times}128$ before inference.
Since a single crop does not span the entire CSD, we apply a sliding-window procedure with horizontal and vertical step sizes of four pixels, pass each crop through the relevant network, and average the resulting probability maps across overlapping crops.
All ML workflows are implemented in Python using the PyTorch framework~\cite{Paszke17-PyTorch}.

%%%%%%%%%%%%%%%%%%%%%%%%%%%%%%%%%%%%%%%%%%%%%%%%%%%%%%%%%%%%%
\subsection{Constant capacitance model}
\label{ssec:constant_cap}
%%%%%%%%%%%%%%%%%%%%%%%%%%%%%%%%%%%%%%%%%%%%%
To perform electrostatic characterization, we use the constant-capacitance formalism (see, e.g., Refs.~\cite{Wiel02-DQD, vanStraaten24-QAR, Gualtieri25-QDS}).
The free energy of a system with $N_d$ hole QDs and $N_g$ gates is
\begin{equation} \label{eq:dot_free_energy}
    U(\vec N) = \frac{e^2}{2} \vec N^T C_{dd}^{-1} \vec N - e\left[C_{dd}^{-1} C_{gd} \vec V_g \right]^T \vec N,
\end{equation}
where $\vec N$ is the vector of dot occupations, $C_{dd}$ is the $N_d \times N_d$ dot–dot capacitance matrix, $C_{gd}$ is the $N_d \times N_g$ gate–dot capacitance matrix, and $\vec V_g$ is the vector of gate voltages.
All $C_{dd}$ and $C_{gd}$ capacitance matrices are in Maxwell form.
Terms independent of $\vec N$ are omitted.

The chemical potential of dot $j$ is the energy required to add one charge to that dot while keeping all others fixed, $\mu_j = U(n_j + 1) - U(n_j)$.
Up to constant offsets, the vector of chemical potentials $\vec\mu$ can be written as
\begin{equation} \label{eq:chemical_potential}
    \vec \mu = e^2\,C_{dd}^{-1} \vec N + \alpha\, \vec V_g,
\end{equation}
where $\alpha$ is the $N_d \times N_g$ lever-arm matrix
\begin{equation} \label{eq:lever_arm}
    \alpha = - e\,C_{dd}^{-1}\,C_{gd}.    
\end{equation}

All relative values of $\alpha$ can be determined from CSDs measured as functions of two gates, $V_{g_1}$ and $V_{g_2}$.
Along a loading line for dot $j$, the corresponding chemical potential is constant.
Differentiating Eq.~(\ref{eq:chemical_potential}) with respect to $V_{g_1}$, we obtain
\begin{equation} \label{eq:slope_1}
    \frac{\alpha_{g_1 : j}}{\alpha_{g_2 : j}} = - \frac{dV_{g_2}}{dV_{g_1}} = - s_j,
\end{equation}
where $\alpha_{g_i : j}$ is the lever arm of gate $i$ on dot $j$ and $s_j$ is the slope of the loading line.
Along an interdot transition between dots $i$ and $j$, the detuning $\varepsilon_{ij} = \mu_i - \mu_j = 0$, which yields
\begin{equation} \label{eq:slope_2}
    s_{ij} = -\frac{\alpha_{g_1:i} - \alpha_{g_1:j}}{\alpha_{g_2:i} - \alpha_{g_2:j}},
\end{equation}
where $s_{ij}$ is the slope of the interdot transition.
The relationships in Eqs.~(\ref{eq:slope_1}) and (\ref{eq:slope_2}) fix the relative lever arms between two gates and two dots.
For dot indices $j \in \{1, 2\}$, and gates $V_{g_1}$ and $V_{g_2}$, we have
\begin{equation} \label{eq:lever arms_from_slopes}
\begin{split}
    \alpha_{g_1 : 2} / \alpha_{g_1 : 1} & = \frac{1 - s_{12} / s_1}{1 - s_{12} / s_2} \\
    \alpha_{g_2 : 1} / \alpha_{g_1 : 1} & = - \frac{1}{s_1} \\
    \alpha_{g_2 : 2} / \alpha_{g_1 : 1} & = \frac{1 - s_{12}/s_1}{s_{12} - s_2}.
\end{split}
\end{equation}
Fixing the overall energy scale such that $\alpha_{g_1 : 1} = 1$, the remaining lever arms follow from the measured slopes.

Once the lever arms are known, we can determine the elements of $C_{dd}$.
The diagonal entries of $C_{dd}^{-1}$ are the charging energies of the dots.
Charging voltages $\Delta V_c^{(j)}$ are obtained from the spacings between successive loading lines for dot $j$, as indicated in Fig.~\ref{fig:mut_charg_planar}(a), and converted to energies by multiplying by the appropriate lever arm.
The off-diagonal entries of $C_{dd}^{-1}$ are the mutual energies, i.e., the shift in the chemical potential of one dot when the occupation of another changes by one.
The corresponding mutual voltages $\Delta V_m$ are also extracted from the CSD and likewise converted to energies via the lever arms.
Finally, given $C_{dd}$ and $\alpha$, we recover the gate–dot capacitances via Eq.~(\ref{eq:lever_arm}),
\begin{equation} \label{eq:cgd}
    C_{gd} = -\frac{1}{e} C_{dd} \, \alpha  .
\end{equation}
Thus, from CSD data, we can reconstruct the complete capacitive model of a quantum dot system, up to an overall energy scale.
To convert $C_{gd}$ and $C_{gd}$ to their non-Maxwell formats, as displayed in Fig.~\ref{fig:mut_charg_planar}, we negate the non-diagonal elements of $C_{dd}$, and we negate $C_{gd}$.
For the $C_{gd}$ results shown in Fig.~\ref{fig:mut_charg_planar} and Fig.~\ref{fig:mut_charg_bilayer}, we set $e = 1$.

%%%%%%%%%%%%%%%%%%%%%%%%%%%%%%%%%%%%%%%%%%%%%%%%%%%%%%%%%%%%%
\subsection{Extracting interdot angles in double-QD CSDs}
\label{ssec:interdot}
%%%%%%%%%%%%%%%%%%%%%%%%%%%%%%%%%%%%%%%%%%%%%%%%%%%%%%%%%%%%%
In this section, we describe how we extract interdot transition angles $\theta_\mathrm{I}$ for double-QD CSDs.
We consider two approaches, referred to as the \textit{direct} and \textit{indirect} methods.

\begin{figure}
    \centering
    \includegraphics{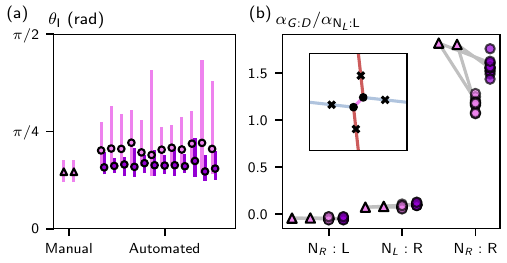}
    \caption{
    \textbf{Comparison direct and indirect method for determining interdot angles $\theta_\mathrm{I}$.}
    (a) Distributions of $\theta_\mathrm{I}$ produced using the direct method (light purple markers) and the indirect method (dark purple markers).
    The two manually-labeled CSD series are marked with triangles. 
    Error bars indicate the 10-90 percentile distributions. 
    (b) Distribution of the resulting terms in the lever arm matrix.
    Each term represents a single series of data, computed from the median transition angles extracted from that series. 
    Inset: Schematic CSD.
    The endpoints of the interdot transition (black dots) are extracted using the centers of L (blue) and R (red) transitions indicated with $\times$-marks and the relevant angles.
    }
    \label{fig:interdot_angles}
\end{figure}

In the direct method, we estimate $\theta_\mathrm{I}$ by applying a Hough transform to the pixels belonging to the detected interdot line.
Results for this approach are shown in Fig.~\ref{fig:interdot_angles}(a) (light purple markers), for both hand-labeled datasets (triangles) and automated datasets (circles).
We observe systematic differences between the hand-labeled and automated $\theta_\mathrm{I}$ values.
Because interdot transitions are typically short and partially obscured by other features, the direct Hough-based estimate is more sensitive to small biases introduced by the ML pixel classification.

%%%%%%%%%%%%%%%%%%%%%%%%%%%%%%%%%%%%%%%%%%%%%
%%%%%%%%%%%%%%%%%%%%%%%%%%%%%%%%%%%%%%%%%%%%%

To mitigate this, we adopt an indirect approach that leverages the left- and right-neighboring loading transitions.
Given a left-loading transition with center $(x_\mathrm{L}, y_\mathrm{L})$ and angle $\theta_\mathrm{L}$, and a right-loading transition with center $(x_\mathrm{R}, y_\mathrm{R})$ and angle $\theta_\mathrm{R}$, we estimate their intersection point $(x_i, y_i)$ as
\begin{equation}
\begin{split}
    x_i &= \left[ 1 - \frac{\tan \theta_\mathrm{R}}{ \tan\theta_\mathrm{L} } \right]^{-1} \left( x_\mathrm{L} - \frac{\tan\theta_\mathrm{R}}{\tan \theta_\mathrm{L}} x_\mathrm{R} + \frac{[y_\mathrm{R} - y_\mathrm{L}]}{\tan \theta_\mathrm{L}} \right) \\
    y_i & = \left( x_i - x_\mathrm{R} \right) \tan \theta_\mathrm{R} + y_\mathrm{R}.
\end{split}
\end{equation}

By computing the intersections of appropriate pairs of left and right transitions, we obtain the endpoints of the interdot segment, schematically indicated in the inset to Fig.~\ref{fig:interdot_angles}(b), from which $\theta_\mathrm{I}$ follows straightforwardly.
Interdot angles obtained with the indirect method, shown as dark purple markers in Fig.~\ref{fig:interdot_angles}(a), agree much more closely with the hand-labeled datasets than do the direct Hough-based estimates.

The impact of this improvement on the extracted lever arms is shown in Fig.~\ref{fig:interdot_angles}(b): lever arms computed using the indirect $\theta_\mathrm{I}$ (dark purple circles) are in markedly better agreement with the manually extracted values (triangles) than those obtained using the direct $\theta_\mathrm{I}$ (light purple circles).
For this reason, the indirect method is used in our automated workflow.

%%%%%%%%%%%%%%%%%%%%%%%%%%%%%%%%%%%%%%%%%%%%%%%%%%%%%%%%%%%%%
\subsection{Computing interlayer mutual voltages}
\label{ssec:interlayer}
\begin{figure}
    \centering
    \includegraphics{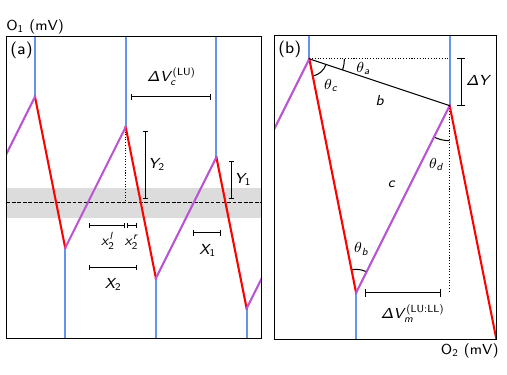}
    \caption{
    \textbf{Extraction of interlayer mutual voltages.} 
    Schematic illustration of a CSD with LU and LL transitions, in the absence of R transitions, highlighting the (a) distances and (b) trigonometric relationships used to compute the interlayer mutual voltage $\Delta V_m^\mathrm{(LU:LL)}$.
    }
    \label{fig:mutual_details}
\end{figure}

%%%%%%%%%%%%%%%%%%%%%%%%%%%%%%%%%%%%%%%%%%%%%%%%%%%%%%%%%%%%%
Here we detail how we compute the interlayer mutual voltage $\Delta V_m^\mathrm{(LU:LL)}$ from the intermediate quantity $\Delta X_m$ introduced in the Results.
For clarity, we first consider the idealized CSD containing only LU, LL, and LI transitions shown schematically in Fig.~\ref{fig:mutual_details}, which corresponds to Fig.~\ref{fig:mut_charg_bilayer}(e).

%%%%%%%%%%%%%%%%%%%%%%%%%%%%%%%%%%%%%%%%%%%%%
%%%%%%%%%%%%%%%%%%%%%%%%%%%%%%%%%%%%%%%%%%%%%

The quantities $X_1$ and $X_2$ in Fig.~\ref{fig:mutual_details}(a) define horizontal spacings between LI and LL transitions in two adjacent cells.
Each $X_i$ can be written as $X_i = x_i^l + x_i^r$, where $x_i^l$ and $x_i^r$ are the contributions from the two sides of the cell, defined as
\begin{equation}
\begin{split}
    x_i^l &= Y_i / \tan \theta_\mathrm{LI} \\
    x_i^r &= Y_i / \tan(\pi - \theta_\mathrm{LL}),
\end{split}
\end{equation}
with $Y_i$ defined as the distance to the intersection point of the LI and LL lines in a cell; see Fig.~\ref{fig:mutual_details}.
This allows us to express the difference $\Delta X_m = X_2 - X_1$ as a change in $Y$
\begin{equation} \label{eq:y_from_x}
    \Delta X_m = \Delta Y\left[ \frac{1}{\tan \theta_\mathrm{LI}} + \frac{1}{\tan(\pi - \theta_\mathrm{LL})} \right].
\end{equation}

We next define the auxiliary angle $\theta_a$ indicated in Fig.~\ref{fig:mutual_details}(b).
Applying the same geometric reasoning as above, we have
\begin{equation}
    \Delta V_c^\mathrm{(LU)} = \frac{\Delta Y}{\tan \theta_a} + \frac{\Delta Y}{\tan \theta_\mathrm{LU}},
\end{equation}
where $\Delta V_c^\mathrm{(LU)}$ is the LU charging voltage.
Solving for $\theta_a$ gives
\begin{equation}
    \theta_a = \tan^{-1} \left[ \Delta Y \left(\Delta V_c^\mathrm{(LU)} - \frac{\Delta Y}{\tan \theta_\mathrm{LU}} \right)^{-1} \right].
\end{equation}
The remaining angles in the triangle of Fig.~\ref{fig:mutual_details}(b) are defined as
$\theta_b = \theta_\mathrm{LL} - \theta_\mathrm{LI}$,
$\theta_c = \pi - \theta_\mathrm{LL} - \theta_a$, and
$\theta_d = \theta_\mathrm{LU} - \theta_\mathrm{LI}$.
Using $\sin \theta_a = \Delta Y / b$, we obtain the side $b$, and by the Law of Sines $c = b \sin(\theta_c) / \sin(\theta_b)$.
A final application of the Law of Sines then yields the mutual voltage
\begin{equation}
    \Delta V_m^\mathrm{(LU:LL)} = c \frac{\sin (\theta_d)}{\sin(\pi - \theta_\mathrm{LU})}.
\end{equation}
In practice we evaluate this expression using median values of $\Delta X_m$, $\Delta V_c^\mathrm{(LU)}$, and the angles $\theta_\mathrm{LU}$, $\theta_\mathrm{LL}$, and $\theta_\mathrm{LI}$ for each series of CSDs.

We now describe how we obtain $\Delta X_m$ from the experimentally acquired CSD.
Let $(x_{\mathrm{LI},j}, y_{\mathrm{LI},j})$ and $\theta_{\mathrm{LI},j}$ denote the center and angle of the LI transition in cell $j$, and $(x_{\mathrm{LL},j}, y_{\mathrm{LL},j})$ and $\theta_{\mathrm{LL},j}$ the corresponding quantities for the LL transition.
The index $j \in \{1,2\}$ labels the two left ($j = 2$) and right ($j = 1$) adjacent cells of the CSD.
The horizontal spacing $X_j$ in cell $j$ is then
\begin{equation} \label{eq:x_mutual}
    X_j = [x_{\mathrm{LL},j} - x_{\mathrm{LI},j}] - \frac{[y_{\mathrm{LL},j} - y_{\mathrm{LI},2}]}{\tan(\theta_{\mathrm{LL},j})} + \frac{[y_{\mathrm{LI},j} - y_{\mathrm{LI},2}]}{\tan(\theta_{\mathrm{LI},j})}.
\end{equation}
From these, we compute $\Delta X_m = X_2 - X_1$ for each pair of adjacent cells, and then use the median $\Delta X_m$ over all valid cell pairs in a series in Eq.~(\ref{eq:y_from_x}).

%%%%%%%%%%%%%%%%%%%%%%%%%%%%%%%%%%%%%%%%%%%%%%%%%%%%%%%%%%%%%
\subsection{Tracking line motion in CSD series}
\label{ssec:tracking_motion}
%%%%%%%%%%%%%%%%%%%%%%%%%%%%%%%%%%%%%%%%%%%%%%%%%%%%%%%%%%%%% 
We use two complementary approaches to estimate the motion of charge transition lines within a series of CSD and calculate the lever arm to the auxiliary gate $V_3$: the Hungarian algorithm~\cite{Kuhn55-HUN} and the Hough transform-based method.

In the first approach, the Hungarian object-tracking algorithm is applied across each series of CSDs to assign all transitions to one of several \textit{tracks}.
The algorithm takes as input all the bounding boxes for transitions of a given type and returns a series of tracks, each ideally corresponding to a single physical transition line over all frames. 
Figure~\ref{fig:leverarm_tracking}(a) shows an example CSD series, with colored boxes indicating three different tracks.
For each track, we then fit the $\mathrm{O}_2$ (or $\mathrm{O}_1$) coordinate as a linear function of $V_3$.
To suppress poorly constrained or misassigned tracks, we discard fits with fewer than $N < 10$ points and fits with $R^2 < 0.7$, where $R^2 = 1 - \sum r_i^2 / (N \sigma^2_{\mathrm{O}_{2(1)}})$, $r_i$ is the residual between the measured coordinate and the linear fit for point $i$, and $\sigma^2_{\mathrm{O}_{2(1)}}$ is the variance of the corresponding coordinate along the track.
For the remaining fits, we take the median slope and use it to estimate the motion of the transition lines with respect to $V_3$.

In the second method, we analyze the same data using a Hough transform.
For each series, we binarize the $V_3$ versus $\mathrm{O}_{2(1)}$ plot into a pixel grid, see Fig.~\ref{fig:leverarm_tracking}(b-ii), and use Eq.~(\ref{eq:hsq}) to compute $H_\mathrm{sq}(\theta)$.
The angle $\theta_\mathrm{max} = \arg\max_\theta H_\mathrm{sq}(\theta)$ characterizes the dominant direction of motion in pixel space.
We note that $H_\mathrm{sq}(\theta)$ is computed in the space of pixels, so $\theta_\mathrm{max}$ does not directly correspond to the slope of $\mathrm{O}_2$ versus $\mathrm{vB}_{TL}$.
Instead, the transition motion $d V_{\mathrm{O}_{2(1)}}^{(D)}/dV_3$ is proportional to $\tan \theta_\mathrm{max}$, where the proportionality constant is given by the unit conversion from pixel back to voltage space.

%%%%%%%%%%%%%%%%%%%%%%%%%%%%%%%%%%%%%%%%%%%%%%%%%%%%%%%%%%%%%
\begin{acknowledgments}
This research was sponsored in part by the U.S. Army Research Office (ARO) under Awards No. W911NF-23-1-0110 and No. W911NF-23-1-0258.
We acknowledge support from the European Union through the IGNITE project with Grant Agreement No. 101069515 and from the Dutch Research Council (NWO) via the National Growth Fund program Quantum Delta NL (Grant No. NGF.1582.22.001) and the Vici ENW (Grant No. VI.C.242.031).

The views and conclusions contained in this paper are those of the authors and should not be interpreted as representing the official policies, either expressed or implied, of the U.S. Government. 
The U.S. Government is authorized to reproduce and distribute reprints for Government purposes, notwithstanding any copyright noted herein. 
Any mention of commercial products is for informational purposes only; it does not imply recommendation or endorsement by the National Institute of Standards and Technology.
\end{acknowledgments}
%%%%%%%%%%%%%%%%%%%%%%%%%%%%%%%%%%%%%%%%%%%%%%%%%%%%%%%%%%%%%

%%%%%%%%%%%%%%%%%%%%%%%%%%%%%%%%%%%%%%%%%%%%%%%%%%%%%%%%%%%%%
%
%%%%%%%%%%%%%%%%%%%%%%%%%%%%%%%%%%%

%%%%%%%%%%%%%%%%%%%%%%%%%%%%%%%%%%%%%%%%%%%%%%%%%%%%%%%%%%%%%
\clearpage % or \newpage to ensure the commands apply to the following page
\clearpage
\setcounter{page}{1}
\setcounter{equation}{0}
\renewcommand{\thepage}{SM-\arabic{page}} 
\renewcommand{\theequation}{SM-\arabic{equation}}
 % Resets the page number counter to 1
\onecolumngrid
\beginsupplement
%%%%%%%%%%%%%%%%%%%%%%%%%%%%%%%%%%%%%%%%%%%%%%%%%%%%%%%%%%%%%

%%%%%%%%%%%%%%%%%%%%%%%%%%%%%%%%%%%%%%%%%%%%%%%%%%%%%%%%%%%%%
\section*{Supplementary Material}
%%%%%%%%%%%%%%%%%%%%%%%%%%%%%%%%%%%%%%%%%%%%%%%%%%%%%%%%%%%%%

%%%%%%%%%%%%%%%%%%%%%%%%%%%%%%%%%%%%%%%%%%%%%%%%%%%%%%%%%%%%%
\section{Processing flow}
\label{sec:model_details}
%%%%%%%%%%%%%%%%%%%%%%%%%%%%%%%%%%%%%%%%%%%%%%%%%%%%%%%%%%%%%
In this section, we provide additional details on the processing flow used to analyze CSDs for both the planar double-QD and the bilayer triple-QD device.
The overall workflow is the same in both cases and proceeds in three steps:
(1) post-processing of the ML model outputs into discrete transition segments;
(2) clustering and tracking of diagonal lines (output of the $\mathcal{M}_{d}$ model) and construction of \textit{cells} that associate vertical and horizontal lines to the neighboring diagonal lines; and
(3) for the bilayer device only, fitting a tri-linear model to the motion of lines from the vertical class to assign fine-grained labels LU, LL, and LI transitions. 
Below, we describe each step and highlight the device-specific parameter choices.

%%%%%%%%%%%%%%%%%%%%%%%%%%%%%%%%%%%%%%%%%%%%%%%%%%%%%%%%%%%%%
\subsection{Post-processing of ML outputs}
\label{ssec:post-processing}
%%%%%%%%%%%%%%%%%%%%%%%%%%%%%%%%%%%%%%%%%%%%%
After passing each CSD through the ML models described in the Methods, we threshold the ML outputs and binarize the corresponding probability maps to obtain discrete transition segments and their respective bounding boxes.
For thresholding, we set $p_{th}=0.2$ across all six ML models.
The specific threshold value is a hyperparameter that can be adjusted if necessary. 

To promote connectivity of thin line segments, we then dilate the binary masks using the OpenCV \texttt{dilate} function with a $2 \times 2$ kernel of ones~\cite{Bradski00-OCV}, for all masks except those produced by the bilayer $\mathcal{M}_h$ model.
For this model, we do not perform additional dilation.
On each binarized and dilated mask, we identify connected components using \texttt{scipy.ndimage.label}~\cite{Virtanen20-SCI}.
Each connected component is treated as a candidate line segment.

For every connected component, we compute the center of mass and a tight axis-aligned bounding box, expanded by one pixel in each direction.
We discard components whose bounding-box diagonal is shorter than $\ell_{th}=10$ pixels.
We also discard lines whose centers lie within $d_{th}=5$ pixels of the CSD boundary to avoid edge effects.
The remaining segments are used as inputs to the geometric analysis and tracking procedures.

%%%%%%%%%%%%%%%%%%%%%%%%%%%%%%%%%%%%%%%%%%%%%%%%%%%%%%%%%%%%%
\subsection{Interdot clustering and cell-based transition association}
\label{ssec:cell-tracking}
%%%%%%%%%%%%%%%%%%%%%%%%%%%%%%%%%%%%%%%%%%%%%
In the second step, we use diagonal segments to define \textit{cells} in which individual horizontal and vertical lines can be tracked across a series of CSD.
This provides a geometric \textit{transition network} framework for relating loading and interdot charge transitions and for tracking individual transitions through time.

For each series, we collect the CSD frame index $f$ and $(x,y)$ pixel coordinates of the center of mass of every detected diagonal segment.
We rescale each coordinate axis to have zero mean and unit variance, and cluster the resulting points using the \texttt{scikit-learn} implementation of the density-based spatial clustering of applications with noise (DBSCAN) algorithm, \texttt{sklearn.cluster.DBSCAN}~\cite{Pedregosa11-SKL}.
For the planar CSD, we cluster the frame index and the $x$-coordinate pair $(f, x)$ pairs with the maximum allowed distance between two pixels set to \texttt{eps}${=}0.65$ and $(f, y)$ pairs with \texttt{eps}${=}0.5$, setting for both \texttt{min\_samples}${=}5$.
For the bilayer CSD, we cluster both $(f, x)$ and $(f, y)$ using \texttt{eps}${=}0.25$ and setting \texttt{min\_samples}${=}10$.
The \texttt{eps} values are optimized through a semistructured grid search over a range centered at the algorithm's default.

Clusters with size less than $5~\%$ of the total point count for a given CSD series are discarded from further analysis.
The surviving clusters are assigned row and column indices along $x$ and $y$, respectively, and serve as nodes of a moving grid within the CSDs.
These clusters are also used to define the relationship between adjacent diagonal lines in the CSD.
For each combination of row and column indices, we take all diagonal segments with these indices, and we fit the positions of these segments as a linear function of $f$.
These fits yield smooth trajectories, which we treat as a \textit{reference grid} that moves throughout the CSD series.

\begin{figure*}[t]
\centering
  \begin{minipage}[c]{0.31\textwidth}
    \caption{
    \textbf{Bounding boxes for right (R) loading transitions (horizontal lines) in bilayer CSDs.}
    (a) A CSD gradient from a series depicted in Fig.~\ref{fig:bilayer_cell_tracking}, highlighting two cells encapsulating horizontal lines.
    The bounding boxes in each cell are the results of clustering and thresholding the output of the $\mathcal{M}_h$ model. 
    The corresponding manually labeled transitions in the same regions are shown in the insets. 
    (b) A CSD from the same series but at a later time. 
    This time, the transition in the right-most cell---which became very weak as the third gate was modified---is missed by the $\mathcal{M}_h$ model. 
    }
    \label{fig:bilayer_right}
  \end{minipage}
  \hfill
  \begin{minipage}[c]{0.68\textwidth}
        \includegraphics{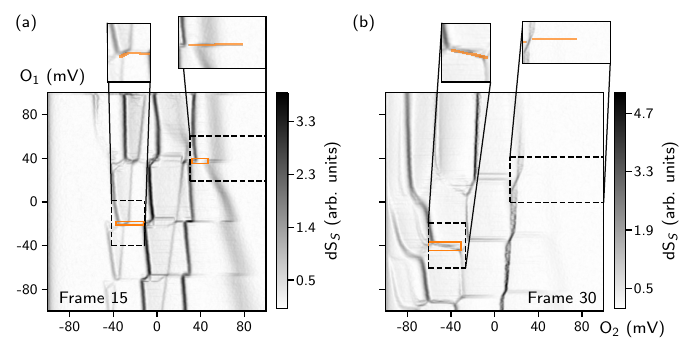}
  \end{minipage}
\end{figure*}

This grid is also used to define the rectangular cell bounds that isolate individual lines (in planar CSDs) and line types (in bilayer CSDs).
For the planar CSDs, cells are chosen so that each contains at most one horizontal or one vertical line.
For the bilayer CSD, special care must be taken to ensure that each cell contains at most one physically distinct vertical transition of each type (LU, LL, or LI).  
In both cases, cells with vertical transitions are oriented between neighboring diagonal lines within the same row.
Cells with vertical lines are expected to be oriented between neighboring diagonal lines in the same column.
Figure~\ref{fig:slope_cluster}(a) shows example cells for one horizontal and one vertical line for a planar CSD highlighted as black dashed boxes.
A series of cells with vertical lines typical for the bilayer CSD is shown in Fig.~\ref{fig:bilayer_cell_tracking}(a) while examples of horizontal lines are shown in Fig.~\ref{fig:bilayer_right}.
Any line whose center lies within a given cell is associated with that cell and its defining diagonal lines.
These cells move with the CSD as the diagonal line locations shift, allowing us to isolate and track trajectories of specific horizontal and vertical lines in time.
Importantly, when a transition is detected within a cell, we can determine its spatial relationship to the other transitions in its neighborhood.

To help define these cells, we first compute the expected width $\mathcal{W}_\mathrm{med}$ and expected height $\mathcal{H}_\mathrm{med}$ of each cell.
The $\mathcal{W}_\mathrm{med}$ is calculated as the median width of all horizontal bounding boxes output by the ML models, and $\mathcal{H}_\mathrm{med}$ as the median height of all vertical bounding boxes output by our ML models.
We also compute $\mathcal{D}_\mathrm{med}$, the median horizontal width of the diagonal segments.

For the vertical lines, we define rectangular cells anchored to the reference grid, with the vertical bounds set by the $y$-coordinates of the diagonal segment directly above and below the cell.
For the bottom row, the lower bound is set to zero; for the top row, the upper bound is set to the size of the CSD. 
If a cell is not in the bottom (top) row but no diagonal segment is detected below (above), we use a default vertical offset of $\mathcal{H}_\mathrm{med}$.
The horizontal extent of each cell is set to $\mathcal{W}_\mathrm{med}$, with the horizontal center determined by the average of the $y$-coordinates of the diagonal segments defining the cell.
For cells in the top row, we add an offset equal to  $\mathcal{D}_\mathrm{med}$ to the horizontal bounds.
For the bilayer device, to ensure we only capture one line of each type in a cell, we make some fine-grained adjustments: we add a constant offset of $-0.3 \times \mathcal{W}_\mathrm{med}$ to the horizontal boundaries for each vertical line cell, except for those in the right-most column, for which we do not adjust the right horizontal bound.

For horizontal lines, we again define rectangular cells anchored to the reference grid, with the horizontal bounds set by the $x$-coordinates of the neighboring diagonal segments to the left and right of the transition.
For transitions on the far right (left) side of a CSD, we set the right (left) bounds in $x$ equal to the size of the CSD (zero).
If either left or right-side diagonal segments are not present, we use a constant cell width of $\mathcal{W}_\mathrm{med}$.
The top (bottom) bound for horizontal line cells is given by the average $y$-coordinate of the diagonal segment immediately to the left and right of the cell with $\pm \mathcal{H}_\mathrm{med}/2$ pixels.
For transitions on the far right (left) side of a CSD, the $y$-coordinate of a single corresponding left (right) diagonal segment is used.

Finally, for transitions within each cell, we perform cell-specific size thresholding.
For vertical lines, we remove detected transitions if their vertical size is less than 20~\% of the vertical cell spacing.
For horizontal lines, we remove detected transitions if their horizontal size is less than 20~\% of the horizontal cell spacing.

%%%%%%%%%%%%%%%%%%%%%%%%%%%%%%%%%%%%%%%%%%%%%
%%%%%%%%%%%%%%%%%%%%%%%%%%%%%%%%%%%%%%%%%%%%%

We note that the resulting cells may not provide a perfect tessellation of the CSDs for any given class of lines.
Rather, some cells partially overlap one or more of the neighboring cells. 
This ensures that the entire bounding box is encapsulated within the respective cell.  
Having defined cells for all horizontal and vertical lines, we can now characterize each of them with respect to the others.

%%%%%%%%%%%%%%%%%%%%%%%%%%%%%%%%%%%%%%%%%%%%%%%%%%%%%%%%%%%%%
\subsection{Tri-linear model fitting for LU/LL/LI (bilayer devices only)}
\label{sssec:model_fit}
%%%%%%%%%%%%%%%%%%%%%%%%%%%%%%%%%%%%%%%%%%%%%
In the bilayer device, the $\mathcal{M}_v$ model detects LU, LL, and LI transitions as a single vertical class.
To assign fine-grained LU, LL, and LI labels, we exploit the characteristic motion of these transitions within each cell across a series of CSDs and fit a tri-linear model, as described in the main text.
Here, we summarize the fitting procedure and the criteria used to select between linear and tri-linear descriptions.
Since not all three transition types are captured in each cell, it is important to consider potential edge cases.
The missing transitions might be due to the quality of the CSD or to the device's underlying state under the local gate voltage configuration. 
To handle these, we adopt the multi-step processing flow described below.

Once all cell boundaries are defined, we collect all transitions belonging to a given cell and consider their $x$-coordinates as a function of frame index $f$, as in Fig.~\ref{fig:bilayer_cell_tracking}.
We then normalize both axes to zero mean and unit variance.

To ensure that the data can be fit reliably, we first discard from further analysis all cell sequences for which the cell is empty in more than 10 frames.
Next, we fit the normalized data to a linear function 
\begin{equation}
    \hat{\tilde x} = \tilde f \tan \theta + \tilde x_0,
\end{equation}
where $\hat{\tilde x}$ is the predicted $x$-coordinate; $\tilde f$ is the frame parameter; $\theta\in(-\pi/2+0.001, 0)$ is the angle between the fitted line and the $x$-axis; and $\tilde x_0\in(-100,100)$ is an offset.
Tilde indicates normalized parameters.
To avoid the effect of outliers, we use an $\mathcal{L}_1$-based per-point error metric 
\begin{equation}\label{eq:lin_err}
\epsilon_\mathrm{lin} = 
\begin{cases}
    |\tilde x_\mathrm{true} - \hat{ \tilde x}| - 0.05, & \text{if } |\tilde x_\mathrm{true} - \hat{ \tilde x } | \geq 0.05, \\
    0, & \text{otherwise},
\end{cases}
\end{equation}
where $\tilde x_\mathrm{true}$ is the normalized coordinate of each data point, and $\epsilon \rightarrow \epsilon - 0.05$ for $\epsilon \geq 0.05$ is a correction used to allow some thickness to the lines.
The total error is defined as $\bar \epsilon_\mathrm{lin} = \sum_i \epsilon_i / N$, with $N$ corresponding to the number of points used in the fit.
To find the best fit, we use the SciPy implementation of the sequential least squares programming (SLSQP) algorithm to minimize $\bar \epsilon_\mathrm{lin}$ over 500 iterations, with initial parameters chosen randomly within the acceptable bounds.
The fits with the lowest total error are used to define the final model.

%%%%%%%%%%%%%%%%%%%%%%%%%%%%%%%%%%%%%%%%%%%%%
\begin{figure*}[t] 
\centering
\hfill
  \begin{minipage}[c]{0.25\textwidth}
        \includegraphics{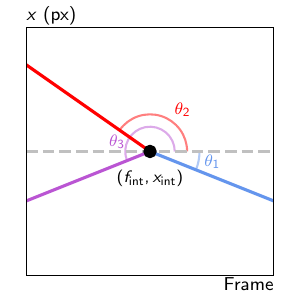}
  \end{minipage}
  \hfill
  \begin{minipage}[c]{0.45\textwidth}
    \caption{
    \textbf{Schematic illustration of the tri-linear model parameters.}
    The tri-linear model has five fit parameters: three angles $\theta_1$, $\theta_2$, and $\theta_3$ defining the slopes of the three lines with respect to the $x$-axis and the intersection point $(\tilde f_\text{int}, \tilde x_\text{int})$ in the normalized (frame index, $x$-coordinate) space. 
    The $x$-axis represents the consecutive frames in the CSD series, while the $y$ axis corresponds to the $x$-coordinates for transitions in the gate-gate space. 
    }
    \label{fig:3lin}
  \end{minipage}
  \hfill
  \begin{minipage}[c]{0.10\textwidth}
  \end{minipage}
\end{figure*}
%%%%%%%%%%%%%%%%%%%%%%%%%%%%%%%%%%%%%%%%%%%%%

Next, we fit the same normalized data to the tri-linear model shown schematically in Fig.~\ref{fig:3lin}.
The per-point error for the tri-linear model $\epsilon_\text{tri}$ is 
\begin{equation}
    \epsilon_\mathrm{tri} = 
    \begin{cases}
        \epsilon^{(1)}_\text{lin}, & \text{if } \tilde x > \tilde x_\text{int} \\
        \min \left( \epsilon^{(2)}_\text{lin}, \epsilon^{(3)}_\text{lin} \right) & \text{otherwise}
    \end{cases}
\end{equation}
where $\epsilon^{(k)}_\text{lin}$ is the per-point linear error associated with branch $k$, for $k={1,2,3}$, given by Eq.~(\ref{eq:lin_err}).
The optimal tri-linear fit is found through a constrained minimization of the average error, $\bar \epsilon_\mathrm{tri}$, using the SLSQP algorithm with 1,000 random initializations.
Cells for which $\bar \epsilon_\mathrm{tri}$ exceeds a threshold $\epsilon_\mathrm{thr} = 0.2$ are discarded for the subsequent analysis.

The empirical constraints for the angles are: $\theta_1 \geq \theta_2 - \pi$, 
$\theta_3 - \theta_2 \geq \pi/20$, and $\theta_3 - \theta_1 \geq \pi$.
To bound the intersection point, we note that the (non-normalized) intersection frame $f_\mathrm{int}$ tends to occur earlier for cells farther right in the CSD (larger $\mathrm{O}_2$) and farther up in the CSD (larger $\mathrm{O}_1$).
Thus, as we explore cells with larger $\mathrm{O}_2$ and $\mathrm{O}_1$ potentials, we compute the intersection point with the smallest $f$ across all cells left of and below the current cell, $(f_\mathrm{int}^\mathrm{min}, x_\mathrm{int}^\mathrm{min})$.
For the subsequent fit, we restrict $f_\mathrm{int} \leq f_\mathrm{int}^\mathrm{min} + 20$ (which we rescale into the appropriate normalized coordinate).

With the errors for both models in hand, we compare the quality of the linear and trilinear fits across all cells. 
Because the tri-linear model has more parameters, we naturally expect $\bar \epsilon_\mathrm{tri} \leq \bar \epsilon_\mathrm{lin}$, even when a linear fit is appropriate.
To capture the crossover between the two descriptions, we use the following condition: if $\bar \epsilon_\mathrm{lin} < \bar \epsilon_\mathrm{thr}$, and $\bar \epsilon_\mathrm{lin} \leq 2 \bar \epsilon_\mathrm{tri} + 0.005$, we select the linear model.

When a linear model is selected, we must assign a single label (LU or LL) to all points in the cell.
Since LI transitions are generally weaker and thus easier to miss for the $\mathcal{M}_{v}$ model, we assume fully LI-only fits are unlikely.
(This makes sense, on physical grounds: interlayer transitions will more weakly affect the sensor QD than charge addition or removal to the same dot.)
When differentiating between LL and LL, cells in the lowest row (smallest $\mathrm{O}_1$) are assigned LL.
For cells not in the lowest row, we consult the model fit for the cell directly below: if its intersection frame $f_\mathrm{int} < 25$ (half the series length), we assign LL; otherwise, LU.
If the cell below is missing, the data is deemed poor quality or not well fit, we fall back to a default rule: cells in the rightmost column or the top row are assigned LU; all others are assigned LL.

Finally, we apply physics-informed quality checks to all accepted tri-linear fits.
First, whenever the number of points classified as LL or LI is fewer than three points, we treat the data as effectively linear.
The same rule applies if the number of LI points is huge (more than four times the number of other points).
In that case, we keep only the LI points and model them linearly, discarding the remainder.
If none of the edge cases apply, we accept the tri-linear fit and use its branch assignments as LU, LL, and LI labels for the points in the cell.

%%%%%%%%%%%%%%%%%%%%%%%%%%%%%%%%%%%%%%%%%%%%%%%%%%%%%%%%%%%%%
\section{Object tracking with Hungarian algorithm}
\label{sec:object_tracking}
%%%%%%%%%%%%%%%%%%%%%%%%%%%%%%%%%%%%%%%%%%%%%%%%%%%%%%%%%%%%%
Tracking the transition lines across different series of CSD allows us to directly measure lever arms between the QD and its corresponding auxiliary gates, as shown in Fig.~\ref{fig:leverarm_tracking}.
The results of this analysis are described in Sec.~\ref{ssec:tracking}.
In this section, we describe the Hungarian-algorithm-based tracking procedure~\cite{Kuhn55-HUN} used to construct transition tracks for all transition types across each CSD series.

\begin{description}[topsep=3pt, itemsep=-2pt, leftmargin=15pt, labelindent=5pt]%after=\vspace{-10pt}, 
    \item[Step 1] Initialize an empty list of active tracks $T$.
    \item[Step 2] Add a track to list $T$ for each transition of given type detected in CSD with frame index $f=0$.
    \item[Step 3] For each frame $f$ in the series do the following:
    \begin{enumerate}[topsep=-2pt, itemsep=-3pt, rightmargin=25pt]
        \item For each track $t \in T$, compute the expected center of the bounding box corresponding to the transition in track $t$ at frame $f$ by fitting the bounding box center coordinates $(x,y)$ of up to the last five detections in that track to a linear function of frame index and extrapolating to $f$. 
        \item Collect the centers of the bounding boxes for all detected transitions of the chosen type in frame $f$.
        \item Construct a cost matrix $\mathcal{C}_{t,d}$ whose rows (index $t$) correspond to tracks, columns (index $d$) indicate detected transitions, and each entry is equal to the distance (in pixels) between the expected and detected transition centers.
        \item Use \texttt{scipy.optimize.linear\_sum\_assignment} to find the assignment of detections to tracks that minimizes the total cost.
        \item Accept a track–detection assignment only if the corresponding distance between predicted and true centers is less than or equal to a maximum allowed distance $d_\mathrm{max}{=}50$~pixels; append accepted detections to their assigned tracks.
        \item Create new tracks for any detections in frame $f$ that are not assigned to an existing track and add them to $T$.
        \item For any track that was not assigned a detection in frame $f$, increment a ``missed'' counter; if a track has no assigned detections for $n_\mathrm{missed}{=}5$ consecutive frames, mark it as frozen and prevent any future detections from being associated with it.  
    \end{enumerate}
\end{description}
The resulting tracks, represented as sequences of {$(f,x,y)$ positions, are then used to estimate the motion of transitions with respect to the third gate and to compute lever arms as described in the main text.
When analyzing the LU and LL, the position sequence is reduced to $(f,x)$, whereas for R it is $(f,y)$; see Methods.
The hyperparameters $d_\mathrm{max}$ and $n_\mathrm{missed}$ can be adjusted depending on the number and resolution of the CSD series.

%%%%%%%%%%%%%%%%%%%%%%%%%%%%%%%%%%%%%%%%%%%%%%%%%%%%%%%%%%%%%
\section{Bilayer CSD Simulations}
\label{sec:csd_fitting}
%%%%%%%%%%%%%%%%%%%%%%%%%%%%%%%%%%%%%%%%%%%%%%%%%%%%%%%%%%%%%
In this section, we provide additional details on the CSD simulations shown in Fig.~\ref{fig:mut_charg_bilayer}(d).
In terms of the charging voltages and lever arms reported in Figs.~\ref{fig:angles_combined} and~\ref{fig:mut_charg_bilayer}, the inverse dot-dot capacitance matrix for the bilayer system is
\begin{equation} \label{eq:cdd_inv_bilayer}
    C_{dd}^{-1} = 
    \begin{pmatrix}
        \Delta V_c^\mathrm{(LU)} & \Delta V_m^\mathrm{(LU:LL)} & \Delta V_m^\mathrm{(LU:R)} \\
        \Delta V_m^\mathrm{(LU:LL)} & 0.5 \times \Delta V_c^\mathrm{(LU)} & \alpha_{\mathrm{O}_2 : \mathrm{LL}} \Delta V_m^\mathrm{(LL : R)} \\
        \Delta V_m^\mathrm{(LU:R)} & \alpha_{\mathrm{O}_2 : \mathrm{LL}} \Delta V_m^\mathrm{(LL : R)} & \alpha_{\mathrm{O}_1 : \mathrm{R}} \Delta V_c^\mathrm{(R)}
    \end{pmatrix},
\end{equation}
where we have dropped the lever arm $\alpha_{\mathrm{O}_2 : \mathrm{LU}}$, since it is defined to be 1.
To populate Eq.~(\ref{eq:cdd_inv_bilayer}), we use the global median-of-medians charging voltages and lever arms across all CSD series.

As described in the main text, we choose the lower-dot charging energy to be defined as $\Delta E_c^\mathrm{(LL)} = 0.5 \times \Delta E_c^\mathrm{(LU)}$, with $\Delta E_c^\mathrm{(LU)} = \Delta V_c^\mathrm{(LU)}$ in our relative energy units.
This choice is consistent with simple constraints.
First, from the device geometry, we expect the lower-layer dots to be larger and less strongly confined than the upper-layer dots.
Second, we know that $\Delta V_c^\mathrm{(LL)} \gtrsim \Delta V_c^\mathrm{(LU)}$, since we do not observe successive LL loading lines in our CSDs.
This implies $\Delta E_c^\mathrm{(LL)} \gtrsim \alpha_{\mathrm{O}_2 : \mathrm{LL}} \Delta V_c^\mathrm{(LL)} \sim 0.2 \times \Delta E_c^\mathrm{(LU)}$.
Within these bounds, the precise numerical choice of $\Delta E_c^\mathrm{(LL)}$ is somewhat arbitrary as long as $\Delta V_c^\mathrm{(LL)} > \Delta V_c^\mathrm{(LU)}$.
This is because changes in $\Delta E_c^\mathrm{(LL)}$ can be largely compensated by adjusting constant offsets along $\mathrm{O}_2$, $\mathrm{O}_1$, and $\mathrm{vB}_{TL}$.

We also include coupling to a third gate, $\mathrm{vB}_{TL}$, in our CSD simulations.
To do so, we extend the lever-arm matrix to three columns
\begin{equation} \label{eq:lever_arm_bilayer}
\alpha = 
\begin{pmatrix}
    1 & \alpha_{\mathrm{O}_1 : \mathrm{LU}} & \alpha_{\mathrm{vB}_{TL} : \mathrm{LU}} 
    \\
    \alpha_{\mathrm{O}_2 : \mathrm{LL}} & \alpha_{\mathrm{O}_1 : \mathrm{LL}} & \alpha_{\mathrm{vB}_{TL} : \mathrm{LL}} 
    \\
    \alpha_{\mathrm{O}_2 : \mathrm{R}} & \alpha_{\mathrm{O}_1 : \mathrm{R}} & 
    \alpha_{\mathrm{vB}_{TL} : \mathrm{R}}
\end{pmatrix}.
\end{equation}
For the $\mathrm{O}_2$ and $\mathrm{O}_1$ lever arms, we use the median-of-medians across all automated datasets; see Fig.~\ref{fig:angles_combined}.
For the $\mathrm{vB}_{TL}$ lever arms, we use the average of the values extracted with the Hough and tracking methods; see Fig.~\ref{fig:leverarm_tracking}.
Given Eqs.~(\ref{eq:cdd_inv_bilayer}) and~(\ref{eq:lever_arm_bilayer}), we obtain $C_{dd}$ and $C_{gd}$ via Eq.~(\ref{eq:cgd}) and use these matrices as input to QArray~\cite{vanStraaten24-QAR, vanStraaten24-QAC} to simulate the bilayer CSDs.

%%%%%%%%%%%%%%%%%%%%%%%%%%%%%%%%%%%%%%%%%%%%%
\begin{figure*}
    \centering
    \includegraphics[width=\textwidth]{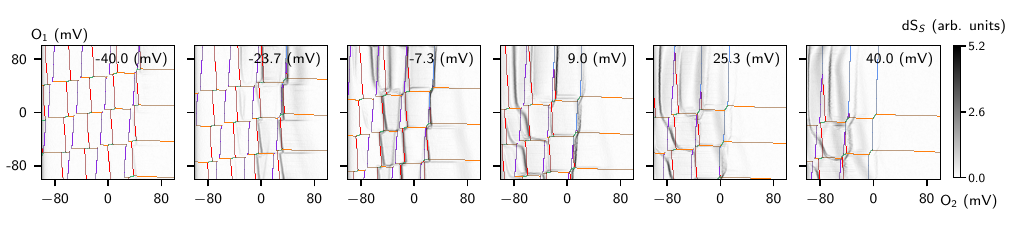}
    \caption{
    \textbf{Additional constant-capacitance simulations of bilayer CSD.}
    Simulated CSDs (colored lines) overlaid on experimental diagrams, as $\mathrm{vB}_{TL}$ is varied from \SI{-40}{\milli\volt} to \SI{40}{\milli\volt}. 
    Constant offsets along $\mathrm{O}_2$, $\mathrm{O}_1$, and $\mathrm{vB}_{TL}$ are used as fitting parameters, chosen to match the simulated and experimental data at $\mathrm{vB}_{TL} = -7.3$~\SI{}{\milli\volt}.
    }
    \label{fig:extended_fit}
\end{figure*}
%%%%%%%%%%%%%%%%%%%%%%%%%%%%%%%%%%%%%%%%%%%%%

In Fig.~\ref{fig:extended_fit}, we show simulated CSDs overlaid on experimental data as $\mathrm{vB}_{TL}$ is varied from \SI{-40}{\milli\volt} to \SI{40}{\milli\volt}.
A CSD for $\mathrm{vB}_{TL} = -7.3$~\si{\milli\volt}, previously shown in Fig.~\ref{fig:mut_charg_bilayer}, is used as the anchor for the fit.
For this frame, we choose constant offsets along $\mathrm{O}_2$, $\mathrm{O}_1$, and $\mathrm{vB}_{TL}$ to align the simulated and experimental diagrams, and we keep these offsets fixed for all other values of $\mathrm{vB}_{TL}$ shown in Fig.~\ref{fig:extended_fit}.
Qualitatively, the simulated and experimental CSDs exhibit very similar behavior: as $\mathrm{vB}_{TL}$ increases, LL and LI lines merge into LU lines that first appear in the upper-right corner of the diagram, and R lines move downward. 
In contrast, LU and LL lines shift left across the CSD.

There are some discrepancies visible, particularly away from the anchor point at
$\mathrm{vB}_{TL} = -7.3$~\si{\milli\volt}, where simulated and experimental lines no longer coincide perfectly.
Several effects contribute to these differences.
First, minor errors in the automatically extracted lever arms and capacitive parameters are amplified as we move farther from the fitting frame.
Second, the experimental device does not strictly adhere to the constant-capacitance assumption: capacitive parameters can drift between series, between CSDs within a series, and even within a single CSD, whereas our simulations use global median-of-medians parameters.
And third, in our capacitive model $\mathrm{vB}_{TL}$ is treated as orthogonal to $\mathrm{O}_2$ and $\mathrm{O}_1$, whereas experimentally $\mathrm{vB}_{TL}$ is defined at an earlier stage of the virtualization procedure~\cite{Rao24-MAViS} and thus is not perfectly compensated in our data.
Nonetheless, the model's ability to reproduce the main qualitative features and trends of the experimental diagrams supports the consistency and utility of our automated extraction procedure.

%%%%%%%%%%%%%%%%%%%%%%%%%%%%%%%%%%%%%%%%%%%%%
\begin{figure*}
    \centering
    \includegraphics[width=\textwidth]{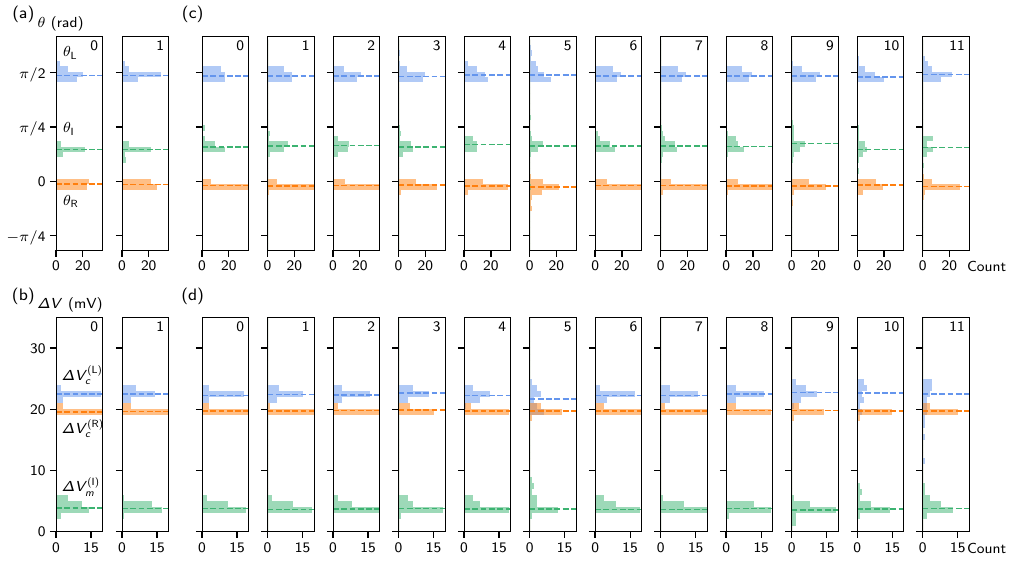}
    \caption{
    \textbf{Histograms for the transition angles and charging voltages for all planar CSD series.}
    (a) Transition angles and (b) charging voltages extracted from the manually labeled CSD series. 
    (c) Transition angles and (d) charging voltages extracted from all automatically characterized CSD series.
    The series number is displayed in the top-right corner.
    Results for series 1 (both manually labeled and automatically characterized) are the same as presented in Figs.~\ref{fig:angles_combined} and \ref{fig:mut_charg_planar} but are included here for completeness.
    }
    \label{fig:mavis_extra_results}
\end{figure*}
%%%%%%%%%%%%%%%%%%%%%%%%%%%%%%%%%%%%%%%%%%%%%

%%%%%%%%%%%%%%%%%%%%%%%%%%%%%%%%%%%%%%%%%%%%%%%%%%%%%%%%%%%%%
\section{Detailed results across all series and all CSDs}
\label{sec:detailed_results}
%%%%%%%%%%%%%%%%%%%%%%%%%%%%%%%%%%%%%%%%%%%%%%%%%%%%%%%%%%%%%
In this section, we include the complete results for all quantities estimated in this work, for both double-dot and bilayer data.
Figure~\ref{fig:mavis_extra_results} shows histograms of transition angles [(a) and (b)] and charging voltages [(c) and (d)] for all double-dot CSD series.
Overall, the histograms are highly consistent across datasets.
Series 6 and 12 exhibit somewhat larger variation, particularly in the charging voltages.
In series 6, this is due to the QD moving from the low- to the high-coupling regime, as discussed in the main text.
Series 12 shows substantial evolution of the CSDs across the series.
In both cases, the broader spread in capacitive parameters is therefore expected.

Next, we provide detailed results for the bilayer device, focusing first on the 3-linear model fits and resulting LU/LL/LI classifications.
Figure~\ref{fig:all_left_lines} shows the fine-grain labels (LU, LL, and LI) for all vertical lines returned by the $\mathcal{M}_{v}$ model for six sample frames from each of the eight bilayer CSD series [(a)--(h)].
In each CSD, black dashed boxes indicate the cells used to isolate left-type transitions, as described in the main text and in Sec.~\ref{ssec:cells}.
The blue, red, and purple bounding boxes correspond to the LU, LL, and LI transition types identified with the tri-linear mode, respectively.
When computing transition angles, lever arms, and related quantities, we retain only the two right-most cells in each row, as indicated in Fig.~\ref{fig:all_left_lines}.

%%%%%%%%%%%%%%%%%%%%%%%%%%%%%%%%%%%%%%%%%%%%%
\begin{figure*}
    \centering
    \includegraphics[width=\textwidth]{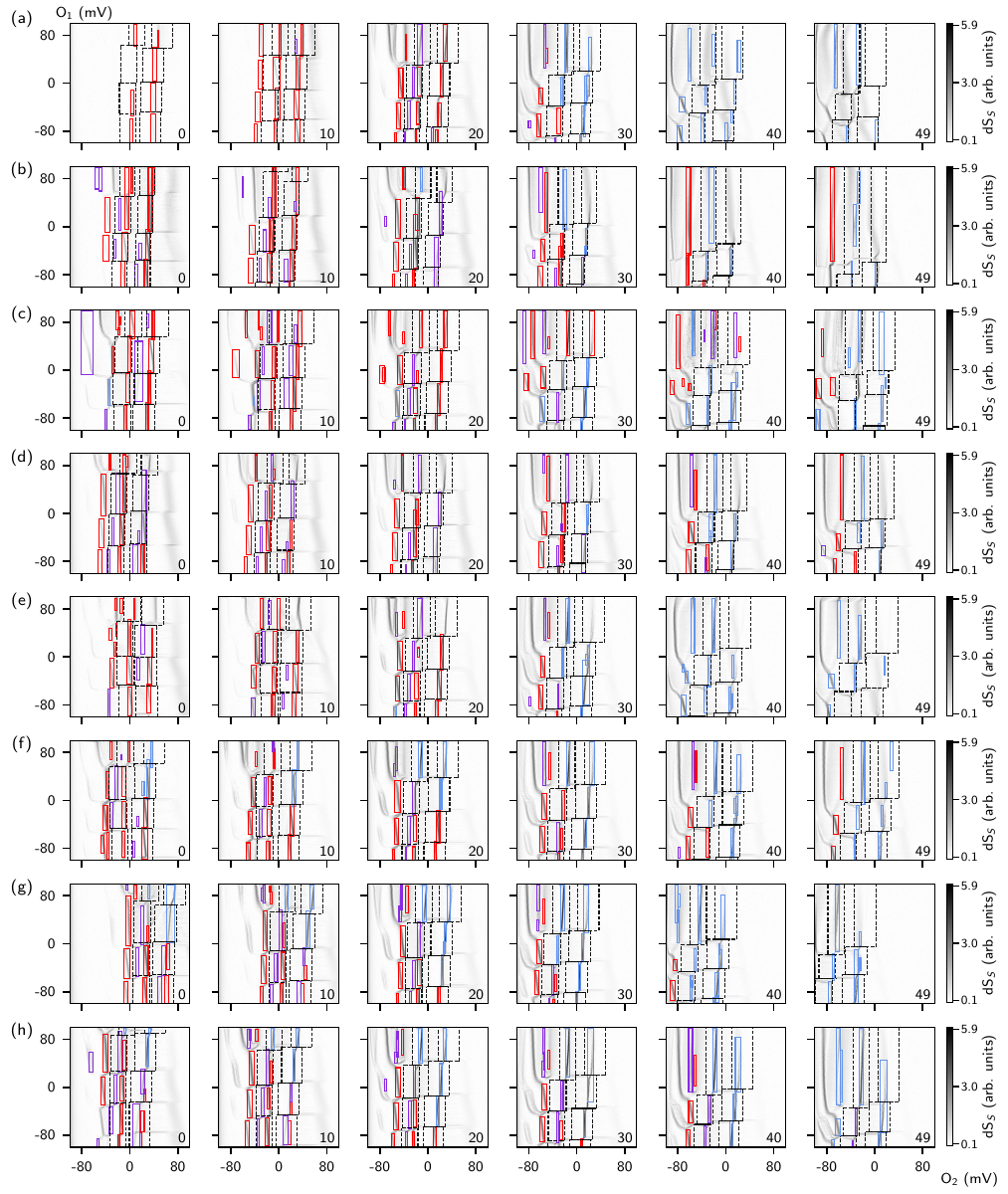}
    \caption{
    \textbf{Results of automatic categorization of the vertical lines in bilayer CSD.}
    All automatically categorized LU (blue), LL (red), and LI (purple) transitions across all eight series of bilayer CSDs [(a)--(h)]. 
    Each series is taken as the center of each CSD is varied along a third gate axis $V_3$, as described in Sec.~\ref{ssec:tracking}.
    The $V_3$ varied in the measurements is (a) $\mathrm{vB}_{TL}$, (b) $\mathrm{vB}_{BL}$, (c) $\mathrm{vB}_{TR}$, (d) $\mathrm{vB}_{BR}$, (e) $\mathrm{S}_{C}$, (f) $\mathrm{S}_R$, (g) $\mathrm{vP}_2$, and (h) $\mathrm{vP}_1$.
    Black-dashed boxes indicate the cells used to track the vertical lines across the CSD series and assign them to one of the physically motivated types. 
    The frame number is displayed in the bottom-right corner of each CSD.
    }
    \label{fig:all_left_lines}
\end{figure*}
%%%%%%%%%%%%%%%%%%%%%%%%%%%%%%%%%%%%%%%%%%%%%

Several fitting and classification errors can occur in the characterization procedure.
First, in the far left region of a CSD, where the coupling between LU and LL dots is much stronger, our fitting procedure occasionally mislabels transitions; this is especially evident in Fig.~\ref{fig:all_left_lines}(c).
Second, some LU transitions split over time into an LL and an LI transition, a behavior that the tri-linear model does not capture.
This effect is most common for transitions farther down and farther left in the CSD series.
For example, in Frame 10 of Fig.~\ref{fig:all_left_lines}(a), the bottom two red transitions in the left-most column should instead be classified as LU (blue).
By Frame 20, two LI lines have appeared, and the LU lines are labeled correctly.
While a more detailed model could account for this behavior, we find it simpler to mitigate both issues by restricting our analysis to the two rightmost cells in each row, where the coupling remains weak and the second failure mode is thus not observed.
Finally, there are occasional fine-grain classification errors that we cannot avoid.
For instance, in the top right-most cells of Frames 30 and 40 in Fig.~\ref{fig:all_left_lines}(c), an LU line is incorrectly identified as an LL/LI pair.
To reduce the impact of unavoidable misclassifications, we consistently use the median as a more robust measure when aggregating quantities from a series of CSDs.
Taken together, these extended results support the consistency and robustness of our automated electrostatic characterization.

Finally, Figs.~\ref{fig:bilayer_angles_extended} and~\ref{fig:bilayer_charging_extended} summarize all quantities extracted from the bilayer CSDs.
Figure~\ref{fig:bilayer_angles_extended} shows angle histograms for both hand-labeled (a) and automated (b) analyses across the eight series of CSDs.
Figure~\ref{fig:bilayer_charging_extended} shows the corresponding charging voltage data: the left-upper and right-dot charging voltages $\Delta V_c^\mathrm{(LU)}$ and $\Delta V_c^\mathrm{(R)}$ [(a) and (b)], the mutual voltages $\Delta V_m^\mathrm{(LU:R)}$ and $\Delta V_m^\mathrm{(LL:R)}$ [(c) and (d)], and the distributions of $\Delta X_m$ used to estimate the interlayer mutual voltage [(e) and (f)].
Across all series, and for all reported quantities, we find good consistency between hand-labeled and automated datasets and between different gate-sweep directions.

%%%%%%%%%%%%%%%%%%%%%%%%%%%%%%%%%%%%%%%%%%%%%
\begin{figure*}
    \centering
    \includegraphics[width=\textwidth]{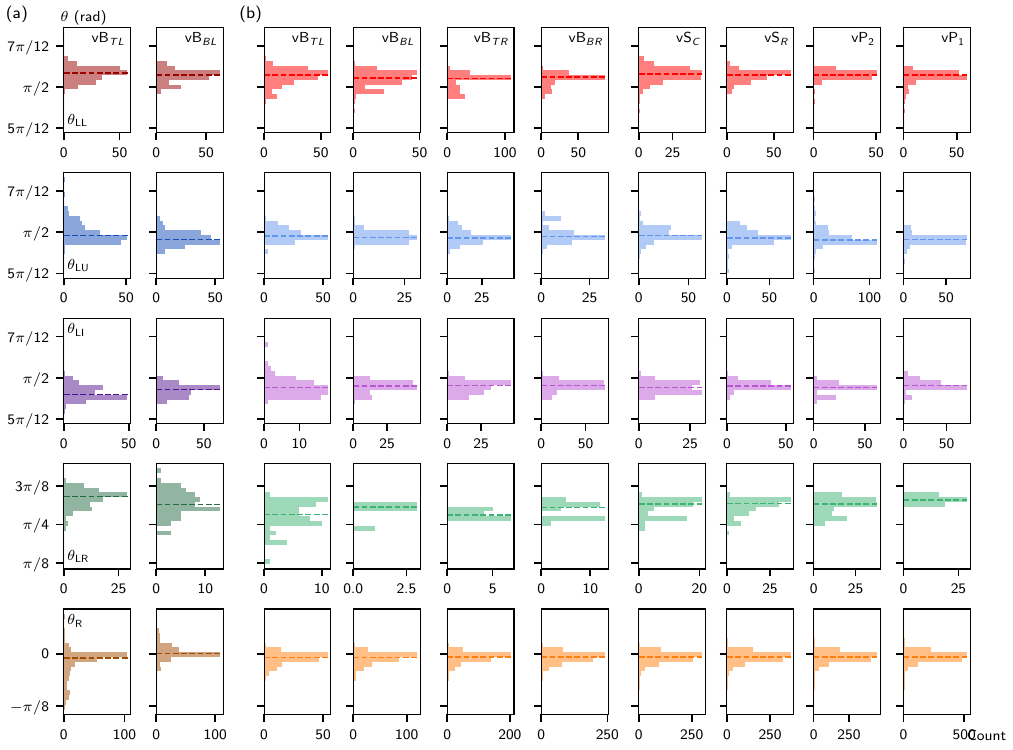}
    \caption{
    \textbf{Histograms of all transition angles extracted from all bilayer CSD series.}
    (a) Distributions of transition angles extracted from manually labeled CSD.
    (b) Distributions of transition angles extracted from automatically characterized CSD.
    Each column represents a single series of CSDs, with the varied gate indicated in the top-right corner of the histograms in the top row.
    Histograms for data in the columns labeled $\mathrm{vB}_{TL}$ are also presented in Fig.~\ref{fig:angles_combined} but are included here for completeness.
    }
    \label{fig:bilayer_angles_extended}
\end{figure*}
%%%%%%%%%%%%%%%%%%%%%%%%%%%%%%%%%%%%%%%%%%%%%

%%%%%%%%%%%%%%%%%%%%%%%%%%%%%%%%%%%%%%%%%%%%%
\begin{figure*}
    \centering
    \includegraphics[width=\textwidth]{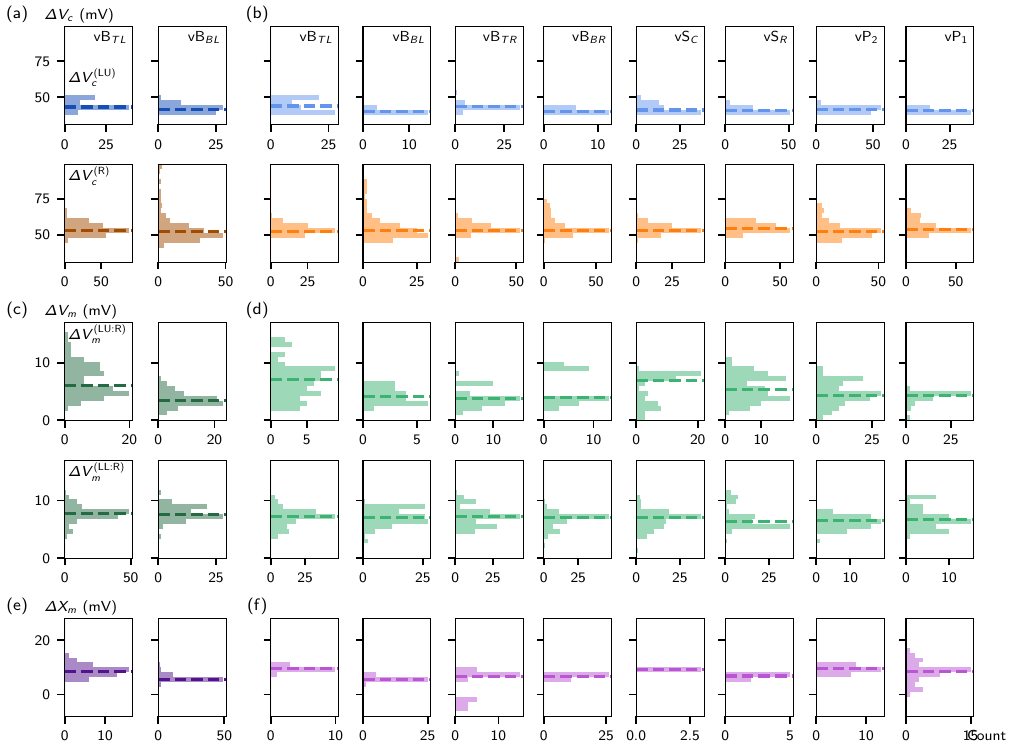}
    \caption{
    \textbf{Histograms of all charging and mutual voltages extracted from all bilayer CSD series.}
    Distributions of the left upper and right dot charging voltages, $\Delta V_c^\mathrm{(LU)}$ and $\Delta V_c^\mathrm{(R)}$ extracted from (a) manually labeled and (b) automatically characterized CSD.
    Distributions of the mutual voltages $\Delta V_m^\mathrm{(LU:R)}$ and $\Delta V_m^\mathrm{(LL:R)}$ extracted from (c) manually labeled and (d) automatically characterized CSD.
    Distributions of the extracted $\Delta X_m$ across (e) manually labeled and (f) automatically characterized CSD.
    Each column represents a single series of CSDs, with the varied gate indicated in the top-right corner of the histograms in the top row.
    }
    \label{fig:bilayer_charging_extended}
\end{figure*}
%%%%%%%%%%%%%%%%%%%%%%%%%%%%%%%%%%%%%%%%%%%%%

%%%%%%%%%%%%%%%%%%%%%%%%%%%%%%%%%%%%%%%%%%%%%%%%%%%%%%%%%%%%%
\end{document}